\documentclass[a4paper,twocolumn,amsmath,amssymb,superscriptaddress,longbibliography,pra, nofootinbib]{revtex4-1}

\usepackage{graphicx}
\usepackage{bm}
\usepackage{braket}
\usepackage{dsfont}
\usepackage[usenames,dvipsnames]{xcolor}
\usepackage{pstricks}
\usepackage[tight]{subfigure}
\usepackage{verbatim}
\usepackage{units}
\usepackage{natbib}
\usepackage{multirow}
\usepackage{enumitem}
\usepackage{mathrsfs}
\usepackage{wasysym}
\usepackage{leftidx}
\usepackage{xspace}
\usepackage{MnSymbol}

\usepackage[normalem]{ulem}  
\usepackage{ulem}

\usepackage[a4paper]{hyperref}
\hypersetup{colorlinks=true,linktoc=all,linkcolor=blue,breaklinks=true,citecolor=blue,urlcolor=blue}

\usepackage[english]{babel}
\AtBeginDocument{%
    \newwrite\bibnotes
    \def\bibnotesext{Notes.bib}
    \immediate\openout\bibnotes=\jobname\bibnotesext
    \immediate\write\bibnotes{@CONTROL{REVTEX41Control}}
    \immediate\write\bibnotes{@CONTROL{%
    apsrev41Control,author="08",editor="1",pages="1",title="0",year="1"}}
     \if@filesw
     \immediate\write\@auxout{\string\citation{apsrev41Control}}%
    \fi
}%

\definecolor{darkgreen}{rgb}{0.0, 0.2, 0.13}
\definecolor{darkorange}{rgb}{1.0, 0.55, 0.0}
\definecolor{shitpink}{rgb}{0.8, 0.35, 0.5}

\newcommand{\RS}[1]{{\color{purple} #1}}

\newcommand{\eg}{e.g.}

\usepackage{hyperref}

\newcommand{\bea}{\begin{eqnarray}}
\newcommand{\eea}{\end{eqnarray}}
\newcommand{\beq}{\begin{equation}}
\newcommand{\eeq}{\end{equation}}
\newcommand{\rme}{\text{e}}

\newcommand{\rmH}{{\mathop{\mathrm{H}}}}
\newcommand{\rmS}{\text{S}}

\newcommand{\rmF}{\text{F}}
\newcommand{\kB}{k_\text{B}}
\newcommand{\kBT}{k_\text{B}T}

\newcommand{\dT}{{\mathop{\delta{T}}}}
\newcommand{\Ag}{{\cal A}}

\begin{document}
	
	
\title{
\vspace*{-1.25cm}
\textnormal{{\small PHYSICAL REVIEW B {\bf 104}, 115430 (2021)}}\\
\vspace*{-0.2cm}
\rule[0.1cm]{18cm}{0.02cm}\\
\vspace*{0.285cm}
Extrinsic thermoelectric response of coherent conductors}

\author{Rafael S\'anchez}
\affiliation{Departamento de F\'isica Te\'orica de la Materia Condensada, Condensed Matter Physics Center (IFIMAC), and Instituto Nicol\'as Cabrera, Universidad Aut\'onoma de Madrid, 28049 Madrid, Spain\looseness=-1}

\author{Cosimo Gorini}
\affiliation{Institut f\"ur Theoretische Physik, Universit\"at Regensburg, 93040 Regensburg, Germany}
\affiliation{Universit\'e Paris-Saclay, CEA, CNRS, SPEC, 91191, Gif-sur-Yvette, France}

\author{Genevi\`eve Fleury}
\affiliation{Universit\'e Paris-Saclay, CEA, CNRS, SPEC, 91191, Gif-sur-Yvette, France}

	
	\begin{abstract}
We investigate the thermoelectric response of a coherent conductor in contact with a scanning probe. Coupling to the probe has the dual effect of allowing for the controlled local injection of heat currents into the system and of inducing interference patterns in the transport coefficients. This is sufficient to generate a multiterminal thermoelectric effect even if the conductor does not effectively break electron-hole symmetry and the tip injects no charge. Considering a simple model for noninteracting electrons, we find a nonlocal thermoelectric response which is modulated by the position of the hot probe tip, and a nonreciprocal longitudinal response which leads to a thermoelectric diode effect. A separate investigation of the effects of dephasing and of quasielastic scattering gives further insights into the different mechanisms involved.
	\end{abstract}

	\maketitle	


\section{Introduction}
\label{sec:intro}

The prominent electronic response to temperature differences in low-dimensional conductors has been discussed for decades, mainly because of their peculiar spectral properties~\cite{hicks:1993,mahan:1996,whitney_most_2014}. Along the same period of time, the field of quantum transport was developed~\cite{Datta1995,ihn_semiconductor_2009,nazarov_quantum_2009}, soon leading to the measurement of the thermoelectric effect in diverse arrangements of zero- and one-dimensional systems~\cite{molenkamp:1990,staring_coulomb_1993, dzurak_observation_1993,dzurak_thermoelectric_1997, small_modulation_2003,llaguno_observation_2004,scheibner_sequential_2007, svensson_lineshape_2012,svensson_nonlinear_2013, thierschmann_diffusion_2013,harzheim_role_2020}, and recently achieving high heat to power efficiencies~\cite{josefsson_quantum_2018}. Quantum coherence in these systems has been suggested to enhance the thermoelectric properties~\cite{finch:2009,bergfield:2009,karlstrom:2011,trocha:2012,gomezsilva:2012,
hershfield:2013}. All these cases are limited by two main characteristics: on one hand, being two terminal measurements, heat is injected longitudinally in the device by the same particles that carry the charge current. It also limits the way heat is injected by increasing the electronic temperature without introducing undesired heat leakage (via the substrate, for instance). On the other hand, the response relies on the energy dependent properties of the nanostructure, in particular the necessity to break the electron-hole symmetry~\cite{benenti:2017}. 

Three-terminal configurations alleviate these limitations by assuming a ``system and gate" geometry: two terminals serve as the conductor where current flows; the third terminal injects on average no charge into the system, and only exchanges heat with it. The response can then be nonlocal when the thermoelectric current is generated in a system at a uniform temperature that includes a region where it interacts with the hot gate~\cite{entin:2010,hotspots}.
A number of configurations have been proposed with a rich variety of properties depending on the nature of the gate, either based on electronic~\cite{hotspots,sothmann:2012,mazza:2014} or bosonic~\cite{entin:2010,entin_three_2012,sothmann:2012epl,ruokola:2012,jiang:2012,hofer:2016prb} interactions, or on complex system-gate couplings~\cite{jordan:2013, bergenfeldt:2014,sanchez:2015qhe, mazza:2015,bosisio_nanowire_2016,sanchez_cooling_2018}, and
experimental realizations have been achieved in systems of quantum dots~\cite{thierschmann:2015,roche:2015,jaliel:2019,dorsch:2020}. 
Applications for thermally driven rectifiers~\cite{sanchez:2011, matthews_thermally_2012,matthews_experimental_2014,thierschmann_thermal_2015,chiraldiode,jiang:2015, rossello:2017,sanchez_single_2017,hwang:2018,donald,acciai:2021} were also reported.
Other kinds of nonlocal thermoelectric effects can be found in hybrid devices~\cite{Cao2015,machon_nonlocal_2013,sanchez_cooling_2018,Hussein2019,Kirsanov2019,Blasi2020a,blasi_nonlocal_2020b,Tan2021}.
However, in all these cases, the thermoelectric response is conditioned
by the system \textit{intrinsic} particle-hole asymmetry --\textit{i.e.}, the system works as a thermoelectric converter even in the absence of the gate. 

\begin{figure}[b]
\includegraphics[width=\linewidth]{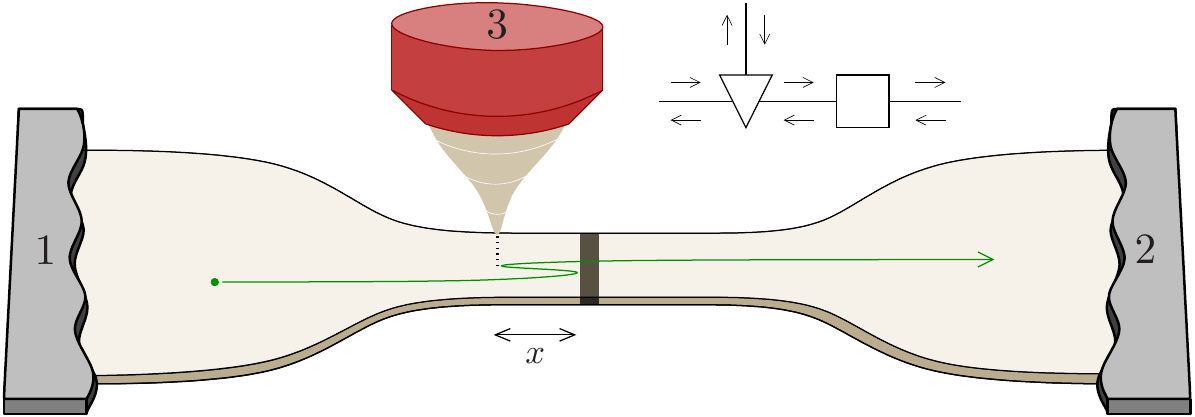}
\caption{One-dimensional conductor with a scattering region (represented by a dark area) coupled to two terminals, 1 and 2, and to a third one via a scanning tip at  $x$. Each terminal, $i$, is characterised by an electrochemical potential $\mu_i$ and a temperature $T_i$. Electron trajectories with multiple internal reflections (one of them is represented by the green arrow) interfere, breaking the conductor electron-hole symmetry. The inset represents the scattering problem schematically. The tip ($\triangledown$) and the scatterer ($\square$) are connected by a single channel.}\label{fig:scheme}
\end{figure}

Here we investigate how a thermoelectric response may be induced into a perfectly symmetric conductor (hereafter called simply {\it conductor}) by a mechanism based on quantum interference only. Transport in the conductor is defined by an energy-independent scattering region (hereafter called the {\it scatterer}) which does not yield any thermoelectric response on its own. In this sense we say the response is extrinsic to the conductor.
As sketched in Fig.~\ref{fig:scheme}, we assume the third terminal to be a scanning tunneling probe~\cite{binnig1987} 
that injects heat but on average no charge into the conductor. Such a heat source can furthermore be actively connected/disconnected. The overall response will hence 
depend on the tip position $x$ with respect to the scatterer. In particular, the tip presence induces interference patterns \cite{hasegawa_direct_1993,crommie_imaging_1993,crommie_confinement_1993} which are measured by the probe electrochemical potential. They are understood in terms of multiple scattering trajectories between the tip and the scatterer~\cite{buttiker:1989}. An example of such a trajectory with multiple internal reflections is shown as a green arrow in Fig.~\ref{fig:scheme}.
The kinetic phase accumulated in these trajectories is sufficient to generate a thermoelectric current, an effect discussed in quantum Hall junctions~\cite{Vannucci2015} and which-path interferometers~\cite{hofer:2015,samuelsson:2017,haack:2019,marchegiani_phase_2020}.

Currents in response to local heating have been recently measured in low dimensional systems via electronic injection~\cite{chen_electron_2012,harzheim:2018,fast:2020,gachter_spatially_2020}, laser illumination~\cite{xu_photo_2010,lemme_gate_2011,zolotavin:2017} or nanoheaters~\cite{mitra_anomalous_2020}. Transport is there dominated by the diffusion of the electron-hole excitations across a potential barrier, what can be interpreted as a nanoscale version of current induced by Landauer blowtorches~\cite{landauer_motion_1988,buttiker_transport_1987,vanKampen_relative_1988}. We assume a simple phenomenological description that intentionally neglects this effect (relying on the intrinsic response of the conductor)  in order to isolate the contribution of quantum interference. We do this by considering a pointlike scatterer with no internal structure.
More realistic descriptions of experimental situations would require the extension of our model to include configurations with more complex scattering properties~\cite{genevieveQPC}. Different types of probes can also be considered~\cite{Park2013,Menges2016,steinacher_scanning_2018,marguerite_imaging_2019}.

Experimentally, 
signatures of oscillations appear overimposed to the intrinsic nonlocal Peltier coefficient of graphene constrictions in Ref.~\cite{harzheim:2018}.
Scanning gate microscopes~\cite{sellier2011,gorini2013,steinacher_scanning_2018} were also recently shown to induce interference fringes in the local thermopower of a quantum point contact~\cite{brun2019}.
Phase-dependent thermopower oscillations were as well reported in hybrid normal-superconducting interferometers~\cite{eom_phase_1998,parsons_reversal_2003}, albeit of very different nature~\cite{jacquod_coherent_2010,kalenkov_large_2017}.

In nonideal configurations, electrons may be affected by events where their phase coherence is lost. A description of the effect of incoherent processes in scattering theory is traditionally done by incorporating additional probe terminals. This approach dates back to the work of Engquist and Anderson~\cite{engquist_definition_1981} and was later refined by B\"uttiker for quantum coherent conductors~\cite{buttiker:1986,buttiker:1988}. Such probes are standardly used to model voltage or temperature measurements, or the effect of inelastic scattering when electrons relax energy in the probe before being reinjected into the conductor. Analogously, quasi-elastic probes have been proposed to describe decoherence~\cite{dejong_semiclassical_1996} when particle currents are conserved at any energy in the probe. However, these probes are invasive in the sense that they introduce additional backscattering in the system: together with phase randomization, they involve momentum relaxation~\cite{Buttiker1991}. This problem is irrelevant in disordered~\cite{dejong_semiclassical_1996} or chiral~\cite{pilgram2006,forster2007} systems, but makes 
such backscattering-inducing probes unsuitable to describe pure dephasing in ballistic conductors. A few works address this issue~\cite{Buttiker1991,Datta1995,Knittel1999,li2002,Golizadeh-Mojarad2007}. We consider the two types of probes (conserving and nonconserving momentum) separately. 
This way, we are first of all able to isolate the effect of pure dephasing on the interference fringes. 
The comparison of pure dephasing and quasi-elastic probe models gives \RS{\sout{useful}} insights into the relevant transport mechanisms. Furthermore, thermometer probes are used to describe (electron-electron) inelastic scattering processes and additionally to measure the effective temperature in the conductor, which is tip position-dependent.

The paper is organized as follows. In Sec.~\ref{sec:scattering} we use scattering theory to describe the transport coefficients, which are analyzed in the linear response in Sec.~\ref{sec:thermoel}. Numerical results are shown in Sec.~\ref{sec:numerics}. Additional probes are included in Sec.~\ref{sec:probes} to describe (momentum conserving and nonconserving) dephasing and temperature probes. Conclusions are discussed in Sec.~\ref{sec:conclusions}.

\section{Scattering theory}
\label{sec:scattering}

A simple and transparent description of the transport problem is given in terms of the scattering formalism for noninteracting electrons~\cite{sivan:1986,streda:1989,butcher:1990,vanHouten:1992}. We adopt here a phenomenological approach where each component of the conductor is described by a minimal scattering matrix imposed by symmetry arguments. This allows us to identify the relevant interference processes involved in the three-terminal thermoelectric response.

\subsection{Scattering matrices}
\label{sec:matrices}

We consider a single-channel one-dimensional conductor connected to two electronic leads, 1 and 2. Transport is assumed to be ballistic except for the presence of a scattering region (represented by a dark stripe in Fig.~\ref{fig:scheme}) at $x=0$ where electrons can be reflected. For concreteness, we will call this region the {\it barrier} in the following. 
We want a bare-bone conductor lacking any intrinsic thermoelectric response.  This is the case if the barrier has no structure, with an energy-independent reflection probability $R$; see \eg, Ref.~\cite{benenti:2017}.
Its scattering matrix can then simply be written as
\begin{eqnarray}\label{Ssq}
\displaystyle
{
S^{\square}=
\left(\begin{array}{cc}
\sqrt{R}\rme^{i\phi} & \sqrt{1{-}R}\rme^{i(\pi+\phi)/2} \\ 
\sqrt{1{-}R}\rme^{i(\pi+\phi)/2} & \sqrt{R}\\
\end{array}  \right),
}
\end{eqnarray}
including the phase $\phi$. 

Electrons are injected into the conductor by a scanning tunneling microscope, see Fig.~\ref{fig:scheme}, which we
model as a pointlike beam splitter~\cite{buttiker:1989} at $x$ coupled to terminal 3. 
Assuming for simplicity that electrons from the tip are injected symmetrically into the other two branches of the beam splitter, we get a scattering matrix $S^\triangledown_{ij}=\sigma_{ij}^\triangledown e^{i(\delta_i+\delta_j)}$, with the orthogonal matrix~\cite{buttiker:1984}:
\begin{eqnarray}\label{Sspl}
\displaystyle
{
\sigma^\triangledown=
\left(\begin{array}{ccc}
-\eta_-/2 & \eta_+/2 & \sqrt{\varepsilon}\\ 
\eta_+/2 & -\eta_-/2 & \sqrt{\varepsilon}\\
\sqrt{\varepsilon} & \sqrt{\varepsilon} & \eta_-{-}1\\
\end{array}  \right),
}
\end{eqnarray}
where $\eta_\pm=1\pm\sqrt{1-2\varepsilon}$. The phases $\delta_i$ preserve the unitarity of $S^\triangledown$. The real parameter $\varepsilon\in[0,1/2]$ represents the tip-conductor coupling. In the limit $\varepsilon=0$, the tip and the conductor are separate systems.
The opposite limit, $\varepsilon=1/2$, describes the case in which the tip has no internal reflection, such that all electrons from the tip terminal are (equally) transmitted into the conductor channels.

The scattering matrix of the whole system is obtained by composing the two matrices $S^{\triangledown}$ and $S^{\square}$ as explained in Appendix~\ref{sec:series}. The local partial densities of states
are different depending on whether the tip is on the left or on the right hand side of the barrier~\cite{gramespacher:1997,gramespacher:1999}, hence leading to different scattering matrices, $S^{-}$ and $S^{+}$, respectively. 

Let us first consider the tip on the left side ($x<0$), as shown in Fig.~\ref{fig:scheme}. The second element of the outgoing waves of the tip is connected to the first element of the incoming waves of the conductor scattering region. Along the way between the tip and the scatter, they accumulate a phase $k|x|$ for wavenumber $k$.
In this case, the transmission probabilities ${\cal T}_{ij}^{\pm}=|S_{ij}^{\pm}|^2$ read
\begin{equation}
\label{eq:transm}
{\cal T}_{12}^{-}{=}\frac{(1{-}R)(\eta_+{-}\varepsilon)}{2{\cal A}},\quad
{\cal T}_{13}^{-}{=}\frac{\varepsilon({\cal A}{+}\zeta)}{\cal A},\quad
{\cal T}_{23}^{-}{=}\frac{(1{-}R)\varepsilon}{{\cal A}}.
\end{equation}
They acquire an oscillatory behaviour due to the interference of trajectories with multiple internal reflections between the tip and the barrier, contained in 
\begin{align}
\label{eq:cA}
{\cal A}&=1+R(\eta_-{-}\varepsilon)/2\pm\sqrt{R}\eta_-\cos\chi\\ 
\label{eq:z}
\zeta&=R(\eta_+{+}\varepsilon)/2\pm\sqrt{R}\eta_+\cos\chi,
\end{align}
with $\chi=\chi_0^-+2k\left|x\right|$ and the phase $\chi_0^-=2\delta_2+\phi$ introduced by the two scatterers. Note that choosing the sign of the last term in the coefficients ${\cal A}$ and $\zeta$ simply adds a phase $\pi$ to $\chi$. In the following, we choose $+$. Such coefficients depend on energy via the momentum of the propagating electron, $k=\sqrt{2m(E-U_0)}/\hbar$, where $U_0$ is the local potential energy, which we assume constant: $U_0=0$, for simplicity.

\begin{figure}[t]
\includegraphics[width=\linewidth]{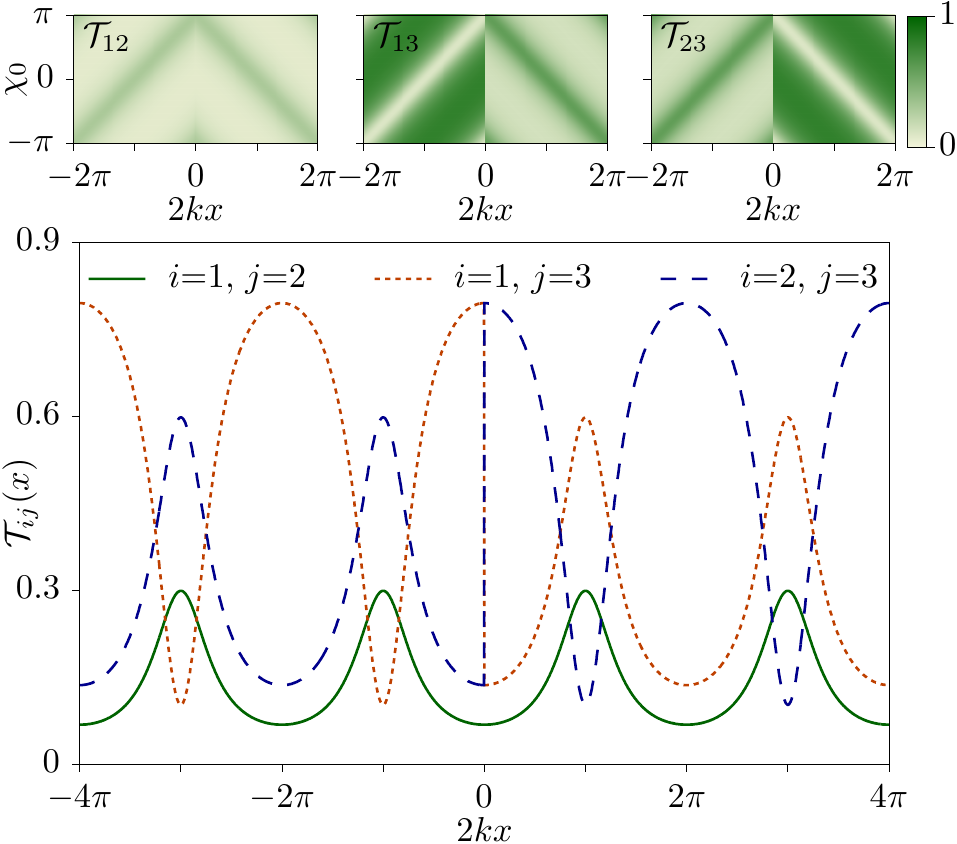}
\caption{Transmission probabilities ${\cal T}_{ij}={\cal T}_{ji}$ between terminals $i$ and $j$ as functions of the tip position and the phase $\chi_0$, for $\varepsilon=R=1/2$. They are periodic in $2kx$ for $x$ not changing sign. When the tip crosses $x=0$, the roles of terminals 1 and 2 are exchanged. The lower panel shows cuts at $\chi_0=0$. 
}\label{fig:transm}
\end{figure}

In the case that $x>0$ (when the tip is on the right side of the barrier), we obtain $S^{+}$ by exchanging 1$\leftrightarrow$2 in the indices of $S^{-}$, cf. Eq.~\eqref{eq:transm}, and replacing $\chi_0^-$ by $\chi_0^+=2\delta_1$. 
For convenience, we consider the symmetric case $\delta_1=\delta_2$ and $\phi=0$, where $\chi_0\equiv\chi_0^+=\chi_0^-$ (except when explicitly stated).
The general expression of the transmission probabilities ${\cal T}_{ij}$ for electrons injected from terminal $j$ to reach terminal $i$ is then obtained by combining
\begin{equation}
\label{eq:Tij}
{\cal T}_{ij}={\cal T}_{ij}^{-}\Theta(-x)+{\cal T}_{ij}^{+}\Theta(x)
\end{equation}
with the Heaviside function $\Theta(x)$.
They are plotted in Fig.~\ref{fig:transm} as functions of the tip position (determining the accumulated phase) and $\chi_0$. Note that the discontinuity of the probabilities ${\cal T}_{13}$ and ${\cal T}_{23}$ at $x=0$ is due to having a point-like scattering region. 

\subsection{Currents}
With all these, we write the particle and heat currents injected from the different terminals:
\begin{align}
\label{eq:ii}
I_i&=\int dE~{\cal I}_i(E)\\
\label{eq:ji}
J_i&=\int dE~(E-\mu_i){\cal I}_i(E),
\end{align}
with the current densities given by 
\beq
{\cal I}_i(E)=\frac{2}{h}\sum_j{\cal T}_{ji}[f_i(E){-}f_j(E)],
\eeq
where $f_i=1/\{1+\exp{[(E-\mu_i)/\kBT_i}]\}$ is the Fermi function of terminal $i$, $h$ and $k_B$ are the Planck and Boltzmann constants, and the factor $2$ takes into account spin degeneracy. The sum is over every terminal in the system. For our particular case they read, for $x<0$:
\begin{align}
\label{eq:i2}
{\cal I}_2&=\frac{2(1{-}R)}{h}\frac{1}{\cal A}\left[\varepsilon(f_2{-}f_3)+\frac{\eta_+{-}\varepsilon}{2}(f_2{-}f_1)\right]\\
\label{eq:i3}
{\cal I}_3&=\frac{2\varepsilon}{h}\frac{1}{\cal A}\left[({\cal A}{+}\zeta)(f_3{-}f_1){+}(1{-}R)(f_3{-}f_2)\right].
\end{align}
The expression for terminal 1 is obtained by particle conservation, ${\cal I}_1=-{\cal I}_2-{\cal I}_3$. 
The currents hence adopt the oscillatory behaviour of the coefficients ${\cal A}$ and $\zeta$~\cite{hasegawa_direct_1993,crommie_imaging_1993,crommie_confinement_1993}. 

The transport problem is solved by assuming probe boundary conditions for the tip. 
Throughout this work, we will assume that the tip is a voltage probe, i.e., its electrochemical potential $\mu_3$ adapts to the condition $I_3=0$, so it does not inject charge in the system on average. The probe is sensitive to the phase accumulated in the conductor and the measured electrochemical potential oscillates with the distance to the barrier~\cite{buttiker:1989}. 

\subsection{Conservation laws and thermodynamics}
\label{sec:conerv}

Under these conditions, charge conservation is expressed only by the conductor terminals
\beq
\label{eq:Icons}
I_1+I_2=0,
\eeq
so the generated particle current is unambiguously defined by one of them. Differently, energy conservation involves all three terminals. In terms of the heat currents, it is written as 
\beq
\label{eq:Jcons}
J_1+J_2+J_3=P
\eeq
and involves the electric power:
\beq
P=-(\mu_2-\mu_1)I_2.
\eeq
We use the convention that $P$ is positive when electrons flow against the chemical potential difference. When this occurs due to heating of terminal $i$, $T_i=T+\delta T_i$, the system works as a converter of heat into useful power, with an efficiency $\eta_i=P/J_i$. Scattering theory respects the second law~\cite{benenti:2017}, so it can be shown that $\eta_i\leq\eta_C$, with the Carnot efficiency $\eta_C=1-T/T_i$.

We anticipate that, when coupling the system to fictitious probes in Sec.~\ref{sec:probes}, the imposed boundary conditions will involve that neither particle nor heat currents are injected. Therefore the above conservation laws for average currents, Eqs.~\eqref{eq:Icons} and \eqref{eq:Jcons}, will not be affected.

\section{Three-terminal thermoelectric response}
\label{sec:thermoel}

An important observation from Eq.~\eqref{eq:i2} is that the thermoelectric response of the conductor vanishes when the tip is not coupled to it. We can easily verify that for $\varepsilon=0$, we have ${\cal A}=1$, and the current reduces to that of an energy-independent two-terminal resistor: $I_2=2(1{-}R)(\mu_2-\mu_1)/h$. Namely, it is independent of the temperatures $T_1$ and $T_2$ of terminals 1 and 2, up to charge accumulation effects in the nonlinear regime~\cite{whitney_nonlinear_2013,sanchez_scattering_2013,meair_scattering_2013} that we neglect here. This ensures that the thermoelectric response discussed below is induced by the presence of the tip only.

To have a finite thermoelectric response, the transmission probabilities ${\cal T}_{ji}$ need to depend on energy. This dependence is introduced by the oscillatory term in ${\cal A}$ and $\zeta$ as soon as $R\neq 0$: if $R=0$ (a perfect conductor), we have ${\cal A}=1$ and $\zeta=0$, and hence $I_2$ and $I_3$ are insensitive to the temperatures $T_1$, $T_2$, and $T_3$. The general problem of solving the integrals in Eqs.~\eqref{eq:ii} and \eqref{eq:ji} with the corresponding boundary conditions is complex and will be solved numerically in Sec.~\ref{sec:numerics}. However, we gain some useful insight by performing a linear response analysis, 
together with a Sommerfeld expansion at low temperature.

\subsection{Linear response analysis}
\label{sec:linear}
Let us assume small electrochemical potential and temperature differences between the different terminals, $\delta\mu_i=\mu_i-\mu\ll\kBT$ and $\delta T_i=T_i-T\ll T$ with respect to the reference electrochemical potential $\mu$ and temperature $T$. The currents are then expanded as
\begin{align}
I_i&=\sum_j\left[G_{ij}(\mu_i-\mu_j)+L_{ij}(T_i-T_j)\right]\\
J_i&=\sum_j\left[M_{ij}(\mu_i-\mu_j)+K_{ij}(T_i-T_j)\right],
\end{align}
where $G_{ij}=2g_{ij}^{(0)}$ and $K_{ij}=2g_{ij}^{(2)}/T$ are the multiterminal particle and heat conductances, while $L_{ij}=2g_{ij}^{(1)}/T$ and $M_{ij}=2g_{ij}^{(1)}$ are the multiterminal thermoelectric coefficients related to the Seebeck and Peltier effects~\cite{butcher:1990}, respectively.
We write them in terms of the integrals
\beq
g_{ij}^{(n)}=
\frac{1}{h}\int{dE}\frac{(E-\mu)^n}{4\kBT}{\cal T}_{ji}(E)\cosh^{-2}\left(\frac{E-\mu}{2\kBT}\right)
\eeq
which make the Onsager reciprocity relations~\cite{Casimir1945} explicit via ${\cal T}_{ji}={\cal T}_{ij}$.

With these response coefficients, we first obtain the probe electrochemical potential by imposing $I_3=0$:
\beq	
\mu_3=\frac{1}{G_{13}+G_{32}}\sum_l[G_{3l}\mu_l+L_{3l}(\delta T_l-\delta T_3)],
\eeq
with the sum limited to the conductor terminals $l=1,2$.
Replacing it in the expression for the current in the conductor and using $\sum_jL_{ij}=0$ (a consequence of charge conservation), we get
\beq
I_2=
\left(G_{21}+\frac{G_{13}G_{23}}{G_{13}+G_{23}}\right)(\mu_2-\mu_1)-\sum_j{\cal L}_{2j}\delta T_j,
\eeq
with the thermoelectric responses
\beq
\label{eq:L2j}
{\cal L}_{2j}=L_{2j}+\frac{G_{23}L_{3j}}{G_{13}+G_{23}}.
\eeq
The current $I_1$ is obtained by replacing 1$\leftrightarrow$2 and one checks that $I_1=-I_2$.   
One can readily see that while the response matrices $G$, $L$, $M$ and $K$  are symmetric, this is not the case, in general, for the thermoelectric responses when not all terminals are equivalent (\eg, one of them is a probe)~\cite{sanchez:2011}: then ${\cal L}_{ij}\neq{\cal L}_{ji}$. 
Note that this implies the possibility of thermoelectric current rectification in the linear regime, in the presence of heat leakage induced by inelastic scattering at the tip terminal~\cite{sanchez:2011}.
In our case, $L_{31}$ and $L_{32}$ will be finite due to interference, as discussed above. The coefficients for heat currents are given in Appendix~\ref{app:linheat}.

We can also verify that ${\cal L}_{2j}=0$ if $\varepsilon=0$ or $R=0,1$, because $g_{ij}^{(1)}=0$ if ${\cal T}_{ji}$ does not depend on energy. 

\subsection{Sommerfeld expansion}
\label{sec:sommerfeld}

A convenient way to picture the relevant processes is to perform a Sommerfeld expansion (see e.g., Ref.~\cite{benenti:2017}) on the linear regime coefficients. For small $\kBT/\mu$, the integrals are evaluated as $g_{ij}^{(0)}={\cal T}_{ji}(\mu)/h$, $g_{ij}^{(1)}=q_{\rm H}{\cal T}_{ji}'(\mu)$ and $g_{ij}^{(2)}=q_{\rm H}{\cal T}_{ji}(\mu)$, where $q_{\rm H}=\pi^2\kB^2T/3h$ is the quantum of thermal conductance~\cite{pendry_quantum_1983}. This is a good approximation when the energy dependence of the transmission coefficient is smooth around $\mu$. In our system, the oscillatory behaviour of ${\cal T}_{ji}$ introduces additional limitations to the applicability of the approximation. For a fixed tip position, they oscillate with a period that depends on $\mu$. Hence, the Sommerfeld expansion stays valid when $\nu_\rmF|x|\ll1/\kBT$ and thermal fluctuations remain negligible (see Appendix \ref{sec:sommlim}). Here, $\nu_\rmF=\sqrt{2m/\mu}/h$ is the 1D density of states at the Fermi energy. For this reason, we will restrict our analysis in this section to tip distances close to $x=0$.

\begin{figure}[t]
\includegraphics[width=\linewidth]{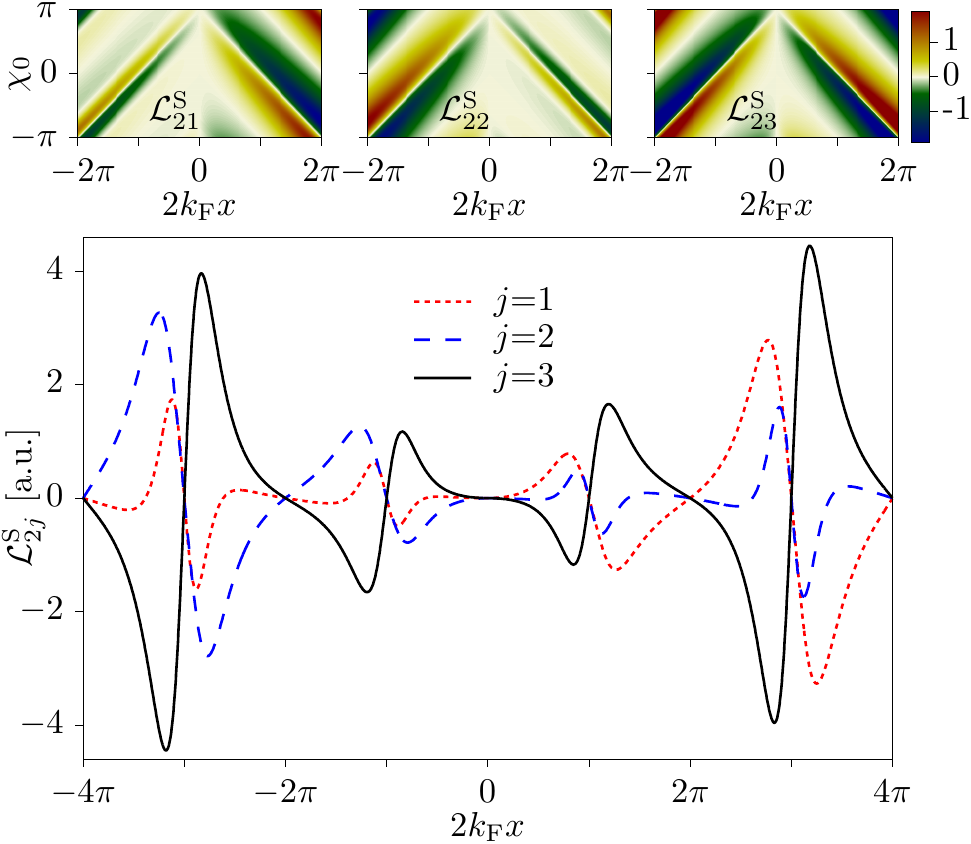}
\caption{Linear response coefficients ${\cal L}_{2j}$ of the current at terminal 2 when increasing the temperature of terminal $j$ by $\delta T_j=\delta T$ as described by a Sommerfeld expansion, for the same parameters as in Fig.~\ref{fig:transm}, with the Fermi wavenumber $k_{\rm F}$. The lower panel shows cuts at $\chi_0=0$ for a doubled range of tip positions.}\label{fig:sommerf}
\end{figure}

The resulting thermoelectric coefficients, ${\cal L}_{2j}^{\rm S}$, are plotted in Fig.~\ref{fig:sommerf}, showing their dependence on the tip position and on the scattering phase $\chi_0$. Clearly, the discontinuity of the transmission probabilities at $x=0$ (see Fig.~\ref{fig:transm}) has an effect in the transport coefficients, which depend on the sign of $x$. However, we note that transport coefficients are reciprocal: ${\cal L}_{ll'}^{\rm S}(x,\chi_0^\pm)={\cal L}_{l'l}^{\rm S}(-x,\chi_0^\mp)$ in the longitudinal terms, and
${\cal L}_{l3}^{\rm S}(x,\chi_0^\pm)={\cal L}_{l'3}^{\rm S}(-x,\chi_0^\mp)$ in the crossed ones, for $l,l'=1,2$ and $l\neq l'$ (remember that ${\cal L}_{1i}^{\rm S}=-{\cal L}_{2i}^{\rm S}$). 

We observe a series of irregular sawtooth-like oscillations as the tip scans the conductor. These are a hallmark of resonances arising from interference. The oscillations shape and sign change depending on which terminal is heated, but they always vanish at $\chi_{\rm F}\equiv2k_{\rm F}x+\chi_0=n\pi$. 
Remarkably, the longitudinal responses have an additional series of nodes when the tip is in between the barrier and the hot terminal (see ${\cal L}_{21}$, for $x<0$, and ${\cal L}_{22}$, for $x>0$). This is possible because the thermoelectric response of the probe (given by its electrochemical potential) has an opposite (and larger) contribution when it is close to the hot terminal, and adds up when it is separated from it by the barrier. We can check this by noticing that $\mu_3\propto L_{3l}\delta T_l\propto{\cal T}_{3l}'(\mu)\delta T_l$ and that ${\cal T}_{13}'$ and ${\cal T}_{23}'$ have opposite signs, cf. Fig.~\ref{fig:transm}. Then, at a fixed tip position the two terms in Eq.~\eqref{eq:L2j} have the same or opposite contributions for $j=1$ and $j=2$, in which case they eventually cancel out. 
Note also that the highest oscillations (in absolute value) correspond to the nonlocal case with a hot tip, while the lowest are those where the hot terminal is on the same side as the tip.

The Sommerfeld expansion also predicts that the generated current increases with the tip-barrier distance, as can be appreciated in Fig.~\ref{fig:sommerf}. This is understood by noticing that all thermoelectric coefficients scale as $L_{ij}^{\rm S}\propto{\cal T}'_{ji}(\mu)\propto\partial_E(k|x|)|_\mu\propto|x|$. In particular, for $x<0$, 
\begin{align}
\label{eq:sommercoeff}
{\cal L}_{2j}^{\rm S}&=q_\rmH\frac{4\pi(1-R)\sqrt{R}}{{\Ag}_\rmF ({\Ag}_\rmF+\zeta_\rmF+1-R)}C_jx\nu_\rmF \sin\chi_\rmF,
\end{align}
where all quantities with a sub-index F are evaluated at the Fermi energy, and
\begin{align}
\label{eq:sommY}
C_1&=2\varepsilon(\eta_+-\varepsilon R)+(\varepsilon-\eta_+)\eta_-({\Ag}_\rmF+\zeta_\rmF+1-R)\nonumber\\
C_2&=(\eta_++\varepsilon)({\Ag}_\rmF+\zeta_\rmF+1-R)-2\varepsilon\\
C_3&=2\varepsilon{\Ag}_\rmF.\nonumber
\end{align}

\begin{figure}[t]
\includegraphics[width=\linewidth]{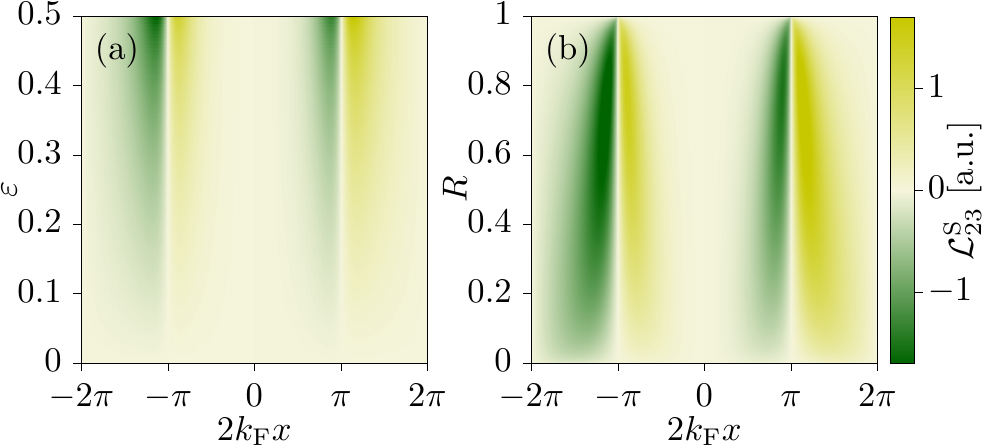}
\caption{Dependence of the nonlocal linear response coefficient ${\cal L}_{23}^{\rm S}$ on (a) the tip coupling $\varepsilon$ (for $R=0.75$) and on (b) the conductor reflection coefficient $R$ (for $\varepsilon=0.5$). In both cases, $\chi_0=0$.}\label{fig:sommerf_epsR}
\end{figure}

Figure~\ref{fig:sommerf_epsR} shows the dependence of the nonlocal term, ${\cal L}_{23}^{\rm S}$, on the coupling parameters $\varepsilon$ and $R$. It increases with the coupling to the tip, $\varepsilon$, and is maximal for $\varepsilon=1/2$, see Fig.~\ref{fig:sommerf_epsR}(a). As expected, the response vanishes both for $R=0$ and $R=1$, when there is no interference, and is maximal around $R\approx3/4$, see Fig.~\ref{fig:sommerf_epsR}(b). 

\section{Numerics}
\label{sec:numerics}

\begin{figure}[t]
\includegraphics[width=0.95\linewidth,clip]{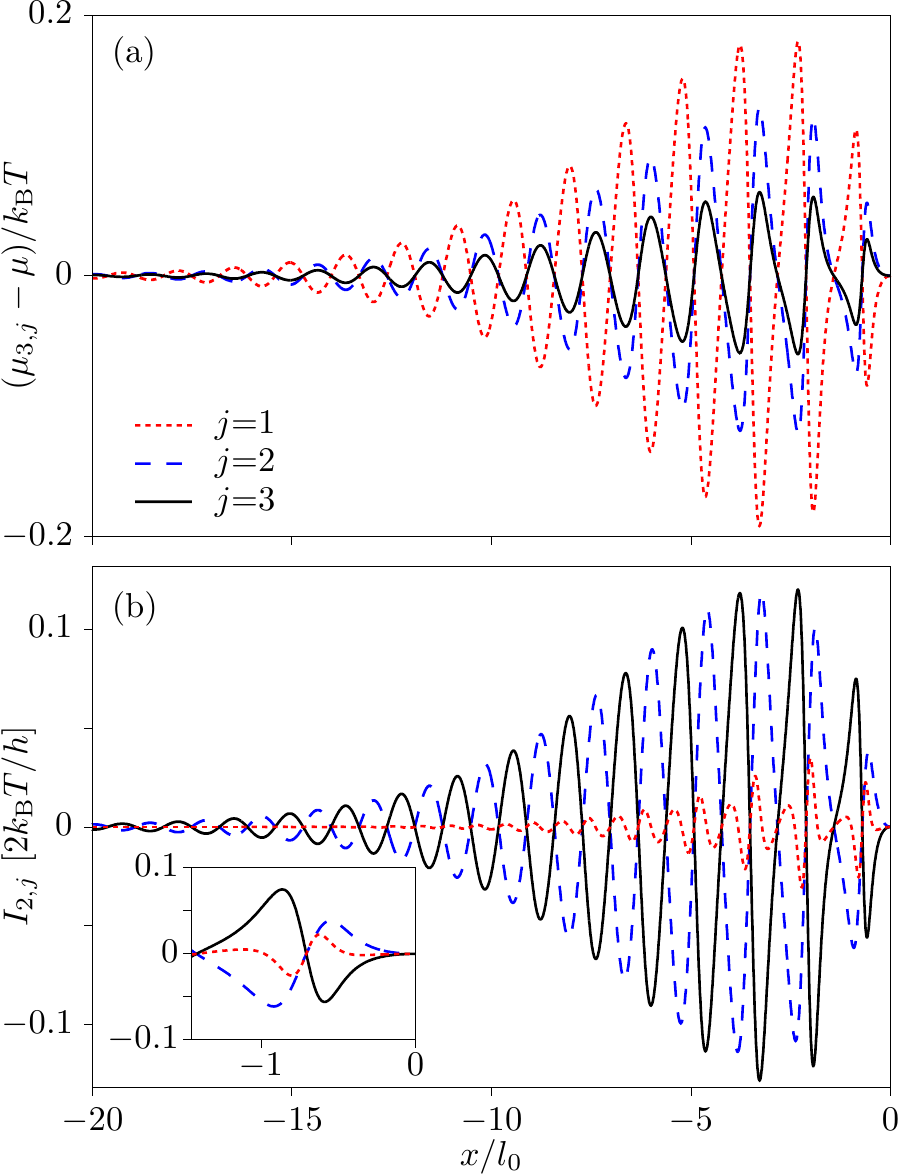}
\caption{\label{fig:muiIi}\small (a) Electrochemical potential measured by the tip, $\mu_{3,j}$, and (b) thermoelectric current, $I_{2,j}$, when the temperature of terminal $j$ is increased by $\delta T=T/2$. The inset in panel (b) shows a zoom on the currents at short distances. Parameters: $\varepsilon=R=0.5$, $\chi_0=0$, $U_0=0$, and $\mu=20\kBT$.}
\end{figure}

Numerically evaluating the integrals in Eqs.~\eqref{eq:ii} and \eqref{eq:ji}, under the condition that $\mu_3$ is such that $I_3=0$, affords a more complete understanding.
We restrict in the following to $x<0$, so charge currents are given by Eqs. \eqref{eq:i2} and \eqref{eq:i3}. We do this for different configurations, depending on which terminal $j$ holds a temperature increase $\dT$ (and assuming $\mu_1=\mu_2=\mu)$. This results in different probe electrochemical potentials, $\mu_{3,j}$ and currents $I_{2,j}$. The result is plotted in Fig.~\ref{fig:muiIi} as a function of the tip position, with $l_0=\hbar/\sqrt{8mk_{\rm B}T}$. The electrochemical potential of the probe oscillates around $\mu$, cf. Fig.~\ref{fig:muiIi}(a). Note that when terminal 1 is hot, the electrochemical potential has the opposite sign with respect to the other cases. It also gives the oscillations with the largest amplitude. This is in contrast with the generated current in Fig.~\ref{fig:muiIi}(b): the smaller the probe response, the largest the current is.

\begin{figure}[t]
\includegraphics[width=0.95\linewidth,clip]{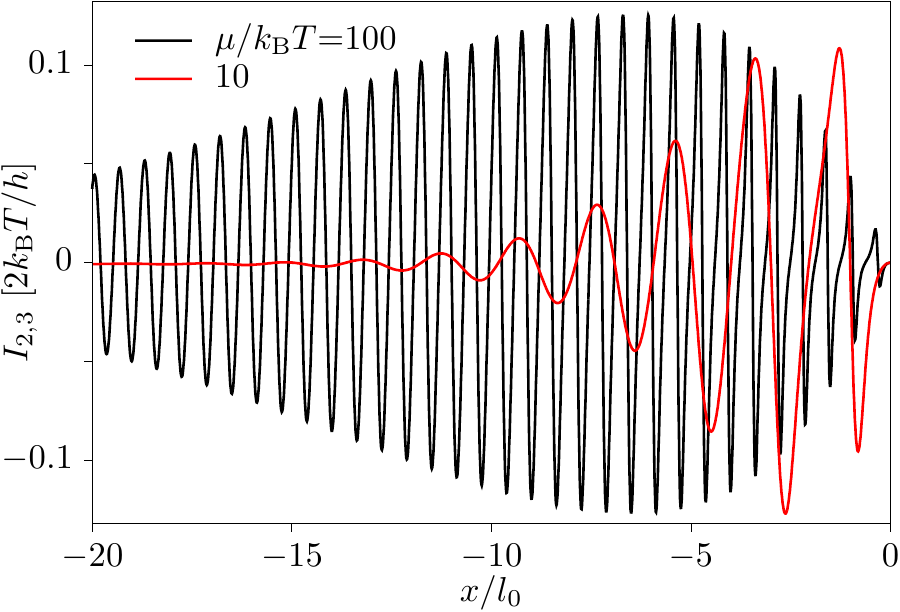}
\caption{\label{fig:thermcurrU}\small Dependence of the nonlocal thermoelectric current ($T_3=T+\dT$) on the electrochemical potential $\mu$, for otherwise the same parameters of Fig.~\ref{fig:muiIi}.}
\end{figure}

The currents in Fig.~\ref{fig:muiIi}(b) reproduce the behaviour expected from the Sommerfeld expansion ($I_{2,j}\approx-{\cal L}_{2,j}^{\rm S}\dT$) at short distances, compare Fig.~\ref{fig:sommerf} with the zoom in the inset of Fig.~\ref{fig:muiIi}(b). 
The same pattern repeats over $x$, with the oscillation amplitudes modulated by the distance. For short distances, the amplitude increases linearly, in agreement with the Sommerfeld expansion. However, there is a crossover to longer distances where the currents decrease. In this regime, the energy dependence of the transmission probabilities around the Fermi energy becomes sensitive to the electrochemical potential $\mu$ through the argument $2k|x|$ of the oscillating term, see Eqs.~\eqref{eq:transm} to \eqref{eq:z}. The energy separation of the oscillations decreases with increasing $|x|$ to a point where they are averaged out in the integration. See Appendix~\ref{sec:sommlim} for a more detailed discussion of this point. This is confirmed by the results shown in Fig.~\ref{fig:thermcurrU}: For larger ratios $\mu/\kBT$, the interference pattern survives at longer distances, and the linear increase for low $x$ can also be more clearly appreciated. For lower $\mu/\kBT$, thermal fluctuations dominate at shorter distances. However, the maximal generated currents are of the same order in both cases.

\subsection{Active tip: Nonlocal thermoelectric current}
\label{sec:nonloc}

Let us focus on the case where the tip is hotter, $T_3=T+\dT$, while $T_1=T_2=T$ and $\mu_1=\mu_2=\mu$. It is interesting to compare this configuration with typical thermocouples where two junctions separate a central hot region (in contact with the heat source) from the rest of the conductor. Those two junctions are furthermore responsible for the separation of electron-hole excitations. Mesoscopic analogs maintain this structure~\cite{jaliel:2019}.
Our case is different, since the conductor has a single barrier. The conductor-tip coupling  provides the mechanism of heat injection from the source (terminal 3), as well as that for thermoelectric current generation. Local thermalization of electrons is not required, i.e., there is no need to define an internal temperature distribution in the conductor (we will however come back to this point in Sec.~\ref{sec:probes}). Transport is hence affected by the nonequilibrium properties of the injected electrons.

In this sense, the thermoelectric properties and the heat source are external to the circuit where current is generated. The tip in a voltage probe configuration injects electron-hole excitations into the system, hence no charge (on average) but heat. In their propagation between the tip and the barrier,
electron and hole quasiparticles acquire different phases, which gives an effect due to the interference of the possible internal reflections. The tip position $x$ hence marks the point where both nonequilibrium excitations and  electron-hole asymmetry are induced. The relevance of the phase coherence will be further explored in Sec.~\ref{sec:probes} by introducing dephasing probes phenomenologically.

\begin{figure}[t]
\includegraphics[width=0.95\linewidth,clip]{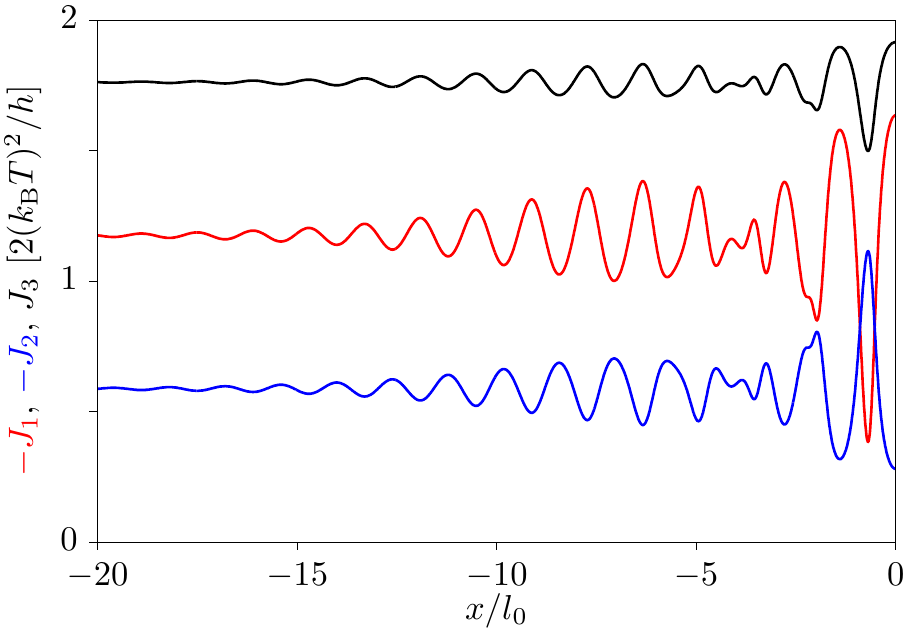}
\caption{\label{fig:Ji}\small Heat currents into the different terminals, $J_i$, when the tip temperature is hot, $T_3=T+\dT$, for the same parameters as in Fig.~\ref{fig:muiIi}.}
\end{figure}

The heat currents due to the coupling to the hot tip are also affected by the interference pattern, as shown in Fig.~\ref{fig:Ji}. Note that, as for  $I_{2,1}$, a double oscillation is apparent for low $x$ that we similarly attribute to the tip probe conditions, see the competition of terms in the linear coefficients in Appendix~\ref{app:linheat}. However, they are always finite. As expected, for the tip being far from the barrier, the heat current is larger in the terminal that is the closest to the tip i.e., $J_1>J_2$ for $x<0$. As the tip approaches the barrier, oscillations appear with opposite phase in terminals 1 and 2 and increase their amplitude to a point that eventually makes the heat current through the barrier larger than that flowing directly to the nearest terminal ($J_2>J_1$, in this case). 

\begin{figure}[t]
\includegraphics[width=\linewidth,clip]{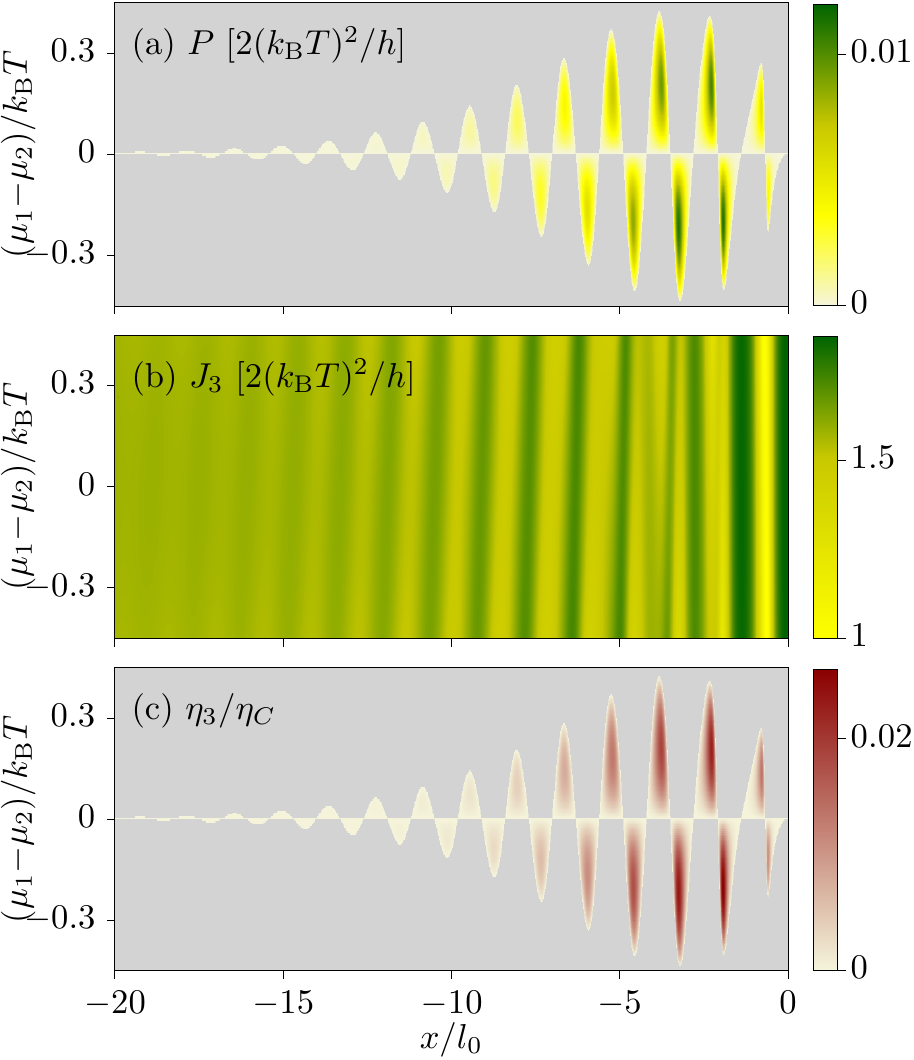}
\caption{\label{fig:Peta}\small Performance of the nonlocal thermoelectric effect. (a) Generated power, (b) injected heat current from the hot tip (with $T_3=T+\delta T$), and (c) efficiency of the heat to power conversion, for $\varepsilon=0.5$ and $R=0.75$. Other parameters are as in Fig.~\ref{fig:muiIi}. The light-grey background in (a) and (c) marks regions where power is dissipated, $P<0$.}
\end{figure}

As there is always one terminal into which electrons flow without being scattered, a large amount of heat will be absorbed by the conductor without contributing to generating a charge current. For this reason, the injected heat current, $J_3$, is expected to be much larger than the power generated in the nonlocal thermoelectric conversion, $P=-(\mu_2-\mu_1)I_2$. We show this in Figs.~\ref{fig:Peta}(a) and \ref{fig:Peta}(b) for the parameters where the linear response coefficient ${\cal L}^S_{23}$ is maximal within Sommerfeld, i.e., with $\varepsilon=0.5$ and $R=0.75$, see Fig.~\ref{fig:sommerf_epsR}(b). Indeed, we find that the efficiency $\eta_3=P/J_3$ is smaller than 2.6\% of the Carnot efficiency, $\eta_C=1-T/T_3$, for the chosen parameters, see Fig.~\ref{fig:Peta}(c). Interestingly, due to the coherent oscillations, the system behaves as a bipolar converter able to produce power for opposite bias voltages but the same temperature configuration, only by shifting the position of the tip.

We speculate that the performance will be improved if the tip is placed between two barriers. Such configuration furthermore introduces an additional interference mode.  However, a detailed investigation of this particular issue is out of the scope of this paper. 

\subsection{Passive tip: Thermoelectric diode}
\label{sec:diode}

A longitudinal thermoelectric effect appears in the case that the hot terminal is in the conductor (either terminal 1 or 2) provided the coupling to the tip is finite. It was discussed above that the response is asymmetric and depends on the position of the tip with respect to the barrier and the heat source [$I_{1,2}(x)=I_{2,1}(-x)\neq I_{2,1}(x)$], cf. Figs.~\ref{fig:sommerf} and \ref{fig:muiIi}. This is possible because energy is relaxed in the voltage probe by means of inelastic scattering~\cite{buttiker:1986}. The probe then induces a thermoelectric rectification effect even in the linear regime~\cite{rossello:2017}. 

The asymmetry does not only affect the magnitude of the currents: To be more precise, we recall that the number of current sign changes with the tip position is doubled when the hot terminal and the tip are on the same side of the barrier; see, e.g., Fig.~\ref{fig:muiIi}(b). Let us restrict for clarity to the case $x<0$ (the case of positive tip positions is obtained by changing terminals 1 and 2 everywhere in the discussion). 
In that case, there exist some positions $x$ of the tip where $I_{2,1}(x)=0$ while $I_{2,2}(x)\neq0$, i.e., charge flows or not through the conductor, depending on which of its terminals is coupled to the heat source, which is very much what one expects of an ideal diode when reversing the sign of the applied voltage. The system then works as an ideal thermoelectric diode. 

\begin{figure}[t]
\includegraphics[width=0.95\linewidth,clip]{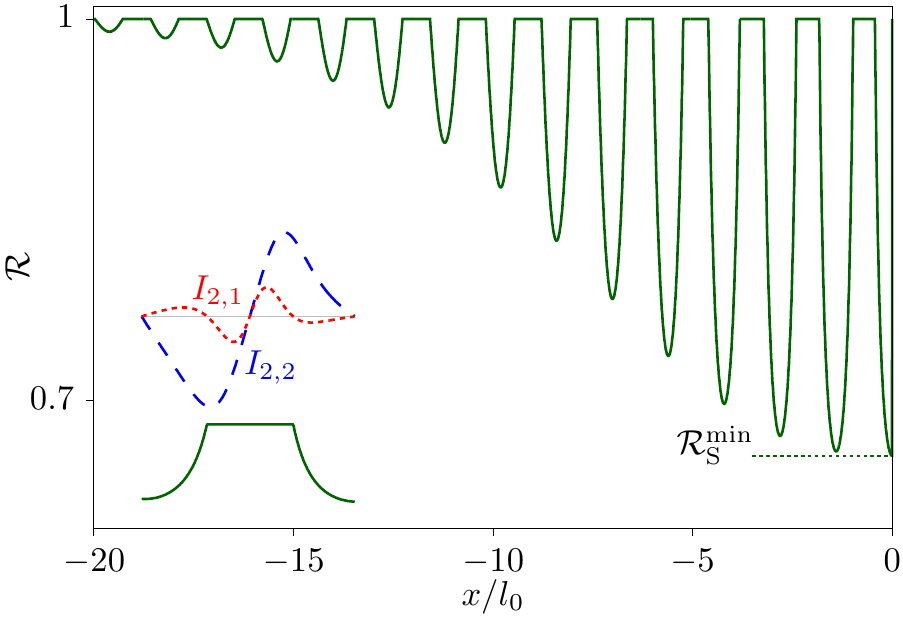}
\caption{\label{fig:rectif}\small Thermoelectric rectification coefficient ${\cal R}$ for the currents plotted in Fig.~\ref{fig:muiIi}(b). A zoom at small $x$ (coinciding with the inset of Fig.~\ref{fig:muiIi}) is shown in the inset for clarity. For short distances, ${\cal R}$ oscillates between 1 and ${\cal R}_\rmS^{\rm min}$, as described by the Sommerfeld expansion in Eq.~\eqref{eq:sommrectif}.}
\end{figure}

To quantify the behaviour of a diode, one compares the forward current $I_\rightarrow=I_{2,1}$ 
measured in terminal 2 when 1 is hot, and the backward one $I_\leftarrow=I_{1,2}$ in the opposite case. Remember we have $I_{1,l}=-I_{2,l}$. 
We parametrize its performance by means of the rectification coefficient:
\beq
{\cal R}=\frac{|I_\rightarrow-I_\leftarrow|}{|I_\rightarrow|+|I_\leftarrow|},
\eeq
which gives ${\cal R}=1$ in the mentioned ideal case and ${\cal R}=0$ if there is no rectification.
As shown in Fig.~\ref{fig:rectif} and its inset, the zeros of $I_{2,1}$ define regions where the thermoelectric current flows in the same direction irrespective of the sign of the temperature gradient along the conductor, resulting in the plateaus with ${\cal R}=1$. Similar features may appear for the heat currents in interacting quantum dot systems because the third terminal acts as a heat sink~\cite{sanchez_single_2017}. However, we emphasize that in our case, the rectified (particle) current is conserved in the conductor.

We note that for tip positions close to the barrier, the rectification coefficient is well described by the results of the Sommerfeld expansion
\beq
\label{eq:sommrectif}
{\cal R}_\rmS=\frac{|C_1+C_2|}{|C_1|+|C_2|},
\eeq
with the coefficients $C_i$ given in Eqs.~\eqref{eq:sommY}.
In this regime, the linear dependence of the currents with $x$ shown in Eq.~\eqref{eq:sommercoeff} makes ${\cal R}_\rmS$ vary between 1 and a constant value (i.e., independent of $x$), marked by ${\cal R}_\rmS^{\rm min}$ in Fig.~\ref{fig:rectif}.
As the current oscillations decrease with the distance (in a regime where the Sommerfeld expansion breaks down), the minima of ${\cal R}$ increase toward 1. In this regime there is a useless strong rectification of tiny currents.

\section{Additional probes}
\label{sec:probes}

The phenomena discussed above arise from quantum interference. In real systems, electrons might however lose phase coherence while propagating along the conductor. It is hence important to explore the robustness of the phenomena to the presence of decoherence. 

\begin{figure}[t]
\includegraphics[width=\linewidth,clip]{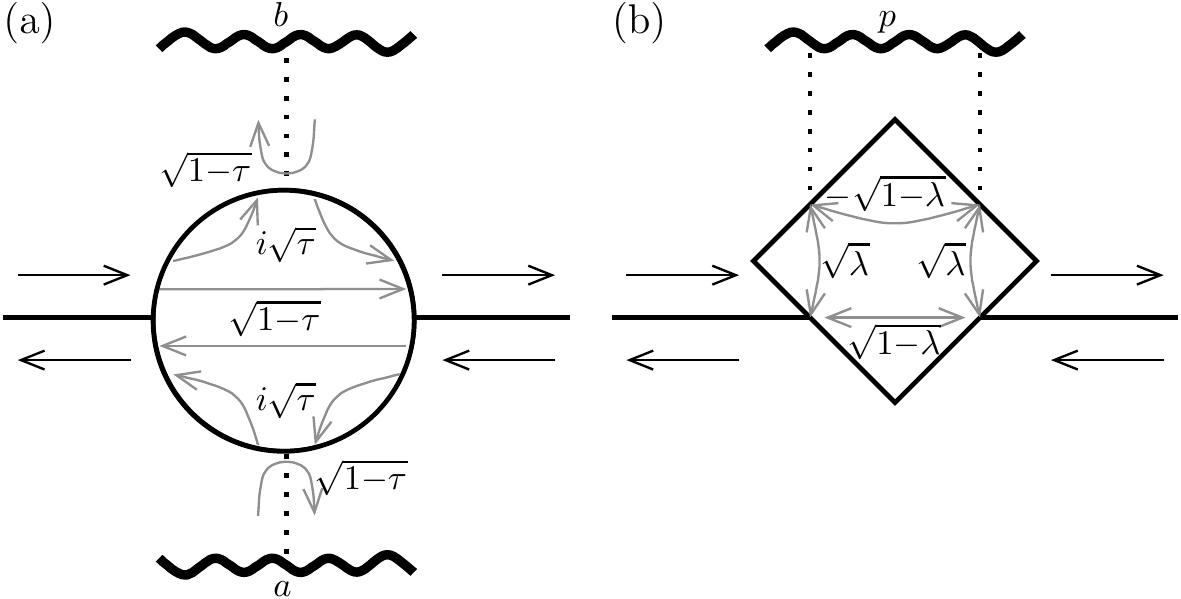}
\caption{\label{fig:Sdeph}\small Representation of the scattering matrices for decoherence processes. (a) $S^{\ocircle}$ describes pure dephasing. Two fictitious probes, $a$ and $b$, are coupled via the dotted channels to the conductor in a chiral way: each of them absorbs and reinjects only left- or right-moving electrons with amplitude $i\sqrt{\tau}$. They have an internal reflection amplitude $\sqrt{1{-}\tau}$. (b) $S^\Diamond$ describes the coupling, with amplitude $\sqrt{\lambda}$, of an invasive probe terminal $p$ inducing backscattering. Depending on the boundary conditions, it is used to model quasielastic or inelastic scattering processes.}
\end{figure}

This is typically done by adding probe terminals to the region between the tip and the barrier. Electrons propagating along the conductor can be absorbed and reinjected by these probes, resulting in phase randomization. The desired properties are phenomenologically described depending on the characteristics of the coupling (given by a scattering matrix) and the boundary conditions imposed to the probe(s). We will consider two kinds of couplings, represented in Fig.~\ref{fig:Sdeph}. They differ in 
whether they introduce phase randomization without backscattering [$S^\ocircle$, cf. Fig.~\ref{fig:Sdeph}(a)], and are hence appropriate for describing pure dephasing, or whether they randomize momentum as well [$S^\Diamond$, cf. Fig.~\ref{fig:Sdeph}(b)]. The latter (which we refer to as invasive in the following) are good for describing decoherence induced by inelastic~\cite{buttiker:1986,buttiker:1988} and quasielastic scattering~\cite{dejong_semiclassical_1996}. Furthermore, they serve as models of thermometers~\cite{engquist_definition_1981,jacquet_temperature_2012,meair:2014,arguelloLuengo_heat_2015,stafford:2016}. 
Experimentally, spatially resolved nonequilibrium distribution~\cite{tikhonov_spatial_2020} and temperature~\cite{halbertal_nanoscale_2016,halbertal_imaging_2017,marguerite_imaging_2019} measurements have been achieved with more involved probes. A simple analytical treatment is however sufficient for our purposes here.

Hereafter, we use the two kinds of probes with different boundary conditions to model separately the effect of pure dephasing, of quasielastic scattering, and of a temperature probe inducing (electron-electron) inelastic scattering.

\subsection{Pure dephasing}
\label{sec:puredeph}
In order to take processes that only affect the phase into account (avoiding additional backscattering), we need a four-channel scattering matrix: two channels [represented by full lines in Fig.~\ref{fig:Sdeph}(a)] correspond to the partitioned channel of the conductor. The other two [dotted lines in Fig.~\ref{fig:Sdeph}(a)] are connected to two fictitious probes, $a$ and $b$, with probability $\tau$. The coupling of the probes to the conductor is chiral~\cite{Buttiker1991}: each of them absorbs electrons from a different ingoing conductor mode and reinjects them in the opposite outgoing mode with the same energy. To mimic pure dephasing, we impose the boundary conditions ${\cal I}_a(E)={\cal I}_b(E)=0$, i.e. charge is conserved at every energy in each probe terminal. This way, after visiting the corresponding probe, where they lose their phase information, electrons continue propagating in the same direction with no energy being exchanged between system and probe. 

We will only consider cases where the dephasing probes are placed between the tip and the barrier. Details on the scattering matrix $S^\ocircle$, as well as how the transmission probabilities are modified and the expressions for the resulting currents are given in Appendix~\ref{sec:probabpd}.

\begin{figure}[t!]
\includegraphics[width=0.95\linewidth,clip]{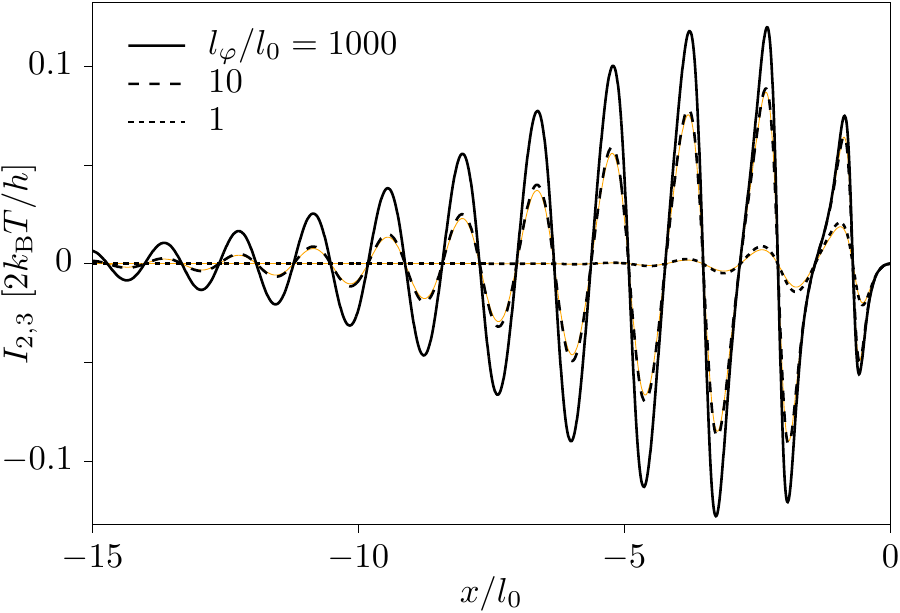}
\caption{\label{fig:crossdeph}\small Effect of pure dephasing on the nonlocal thermoelectric current ($T_3=T+\delta T$), parametrized  by the phase coherence length, $l_\varphi$, defined in Eq.~\eqref{lpdeph}. 
Parameters are as in Fig.~\ref{fig:muiIi}, whose curve $I_{2,3}$ is the limit $l_\varphi\rightarrow\infty$ of this case. As there, $\delta T=T/2$. The cases with decoherence due to a (non momentum conserving) quasielastic scattering for the same values of the length defined in Eq.~\eqref{ldeph} are shown in orange, for comparison.}
\end{figure}

The probe conditions ${\cal I}_a(E)={\cal I}_b(E)=0$ are satisfied only if the probes acquire a nonequilibrium distribution, as discussed in Appendix~\ref{sec:probabpd}, see Eq.~\eqref{eq:purdephdistr}.
The resulting nonlocal current is plotted in Fig.~\ref{fig:crossdeph}. 
We get a more physical insight of the effect of dephasing by introducing a phenomenological dephasing length defined as
\beq
\label{lpdeph}
\tau(x)=1-e^{-|x|/l_\varphi}.
\eeq
That is, the phase coherence of the injected electrons is fully lost before any internal reflection occurs if the tip and the barrier are far enough ($|x|\gg l_\varphi$), as they will likely be absorbed by terminals $a$ or $b$. 
When the tip distance is of the order of the dephasing length, the loss of phase coherence between tip and barrier will only be partial.
This is shown for various $l_\varphi$ in Fig.~\ref{fig:crossdeph}. The current vanishes for tip positions larger than $l_\varphi$, confirming the importance of the interference effect for having a nonlocal thermoelectric response. Conductors with a short coherence length will generate current only for tips in the close vicinity of the barrier.

\begin{figure}[t]
\includegraphics[width=0.95\linewidth,clip]{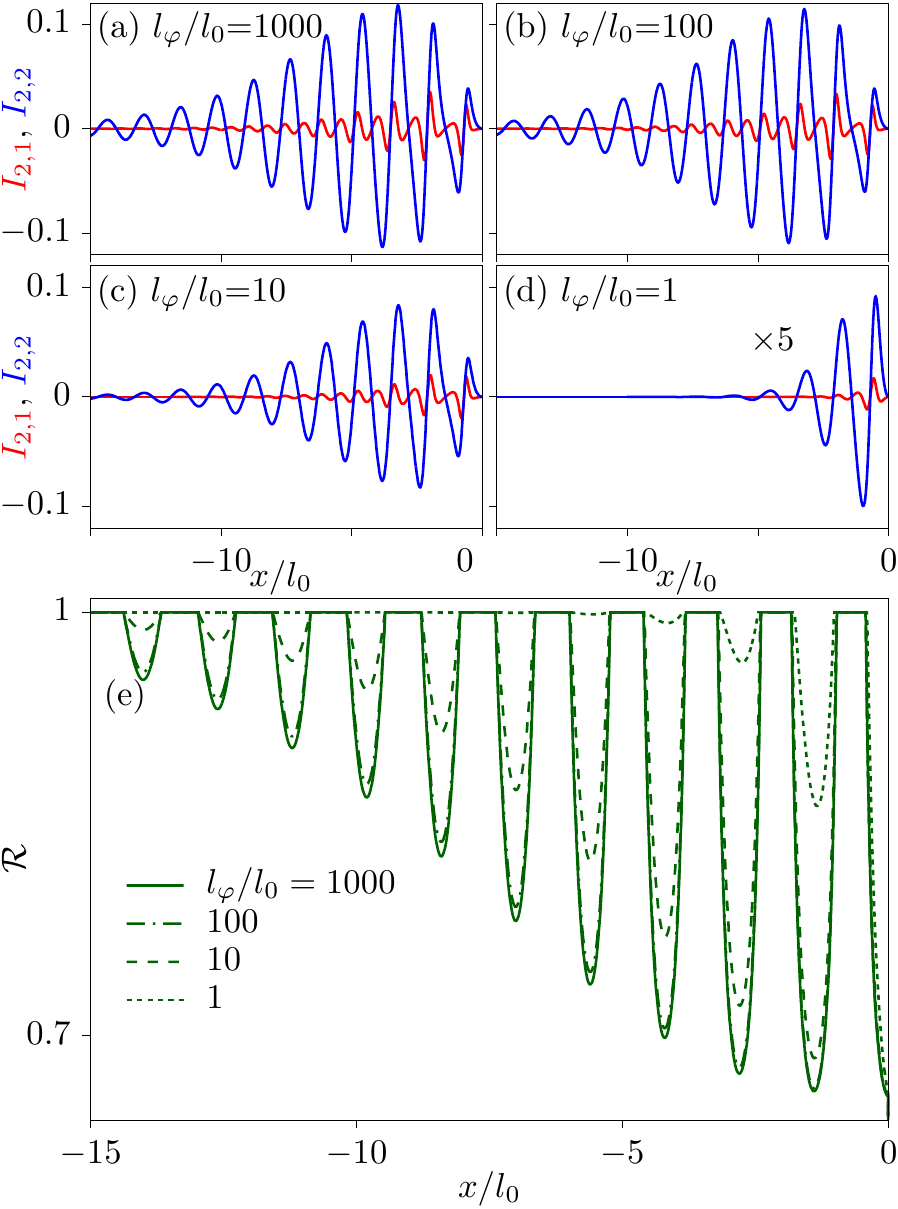}
\caption{\label{fig:longpuredeph}\small Dependence of (a)-(d) the longitudinal currents $I_{2,1}$ and $I_{2,2}$ and (e) the thermoelectric rectification coefficient ${\cal R}(x)$ on the phase coherence length due to pure dephasing, $l_\varphi$. All other parameters are as in Figs.~\ref{fig:muiIi} and \ref{fig:rectif}, which correspond to taking $l_\varphi\rightarrow\infty$.}
\end{figure}

The longitudinal currents are affected by dephasing in a similar way, as shown in Fig.~\ref{fig:longpuredeph}. As the coupling to the probes increases, the amplitude of the oscillations decreases. As the probes respect the electronic propagation, the different features of the currents $I_{\rightarrow}$ and $I_{\leftarrow}$ that resulted in the rectification properties discussed in Sec.~\ref{sec:diode} are maintained, see Fig.~\ref{fig:longpuredeph}(a)-(d). As a consequence, we still find regions where the two currents have the same sign (and give ${\cal R}=1$). Furthermore, the rectification coefficient increases when the dephasing length decreases, see Fig.~\ref{fig:longpuredeph}(e), with the drawback that the rectified currents become small.

\subsection{Quasielastic scattering}
\label{sec:dissprobe}

Dephasing and/or decoherence can also be caused by processes that randomize both phase and momentum of the carriers. A minimal description of these processes is given by connecting the conductor to a single probe terminal, which we will label $p$, where electrons relax before being reinjected into the conductor~\cite{buttiker:1986}, see Fig.~\ref{fig:Sdeph}(b). The coupling probability is in this case $\lambda$.
Processes of different microscopic origins can be mimicked by imposing appropriate boundary conditions at the probe.  We now consider elastic scattering, and discuss inelastic processes in Sec.~\ref{sec:thermometer}.
We give details of the probe scattering matrix $S^\Diamond$ and the corresponding transmission probabilities ${\cal T}_{\alpha\beta}^{\lambda}$, as well as the expression for the resulting currents in Appendix~\ref{app:sdeph}.

The fictitious probe terminal $p$ describes quasielastic dephasing by again imposing energy-resolved boundary conditions, ${\cal I}_p(E)=0$: the probe absorbs and reinjects electrons without changing their energy, but randomizing their phase and momentum~\cite{dejong_semiclassical_1996}. As a result the probe acquires a nonequilibrium distribution, cf. Eq.~\eqref{eq:f4}.

A phenomenological phase coherence length $l_\varphi$ can also be defined in this case as
\beq
\label{ldeph}
\lambda(x)=1-e^{-|x|/l_\varphi}.
\eeq
Its effect on the nonlocal response is shown in Fig.~\ref{fig:crossdeph}, being almost indistinguishable from the case of pure dephasing, especially for short coherence lengths. However, we observe that the amplitude of the oscillations is most effectively reduced by the quasielastic scattering processes, due to the additional relaxation of momentum. 

\begin{figure}[t]
\includegraphics[width=0.95\linewidth,clip]{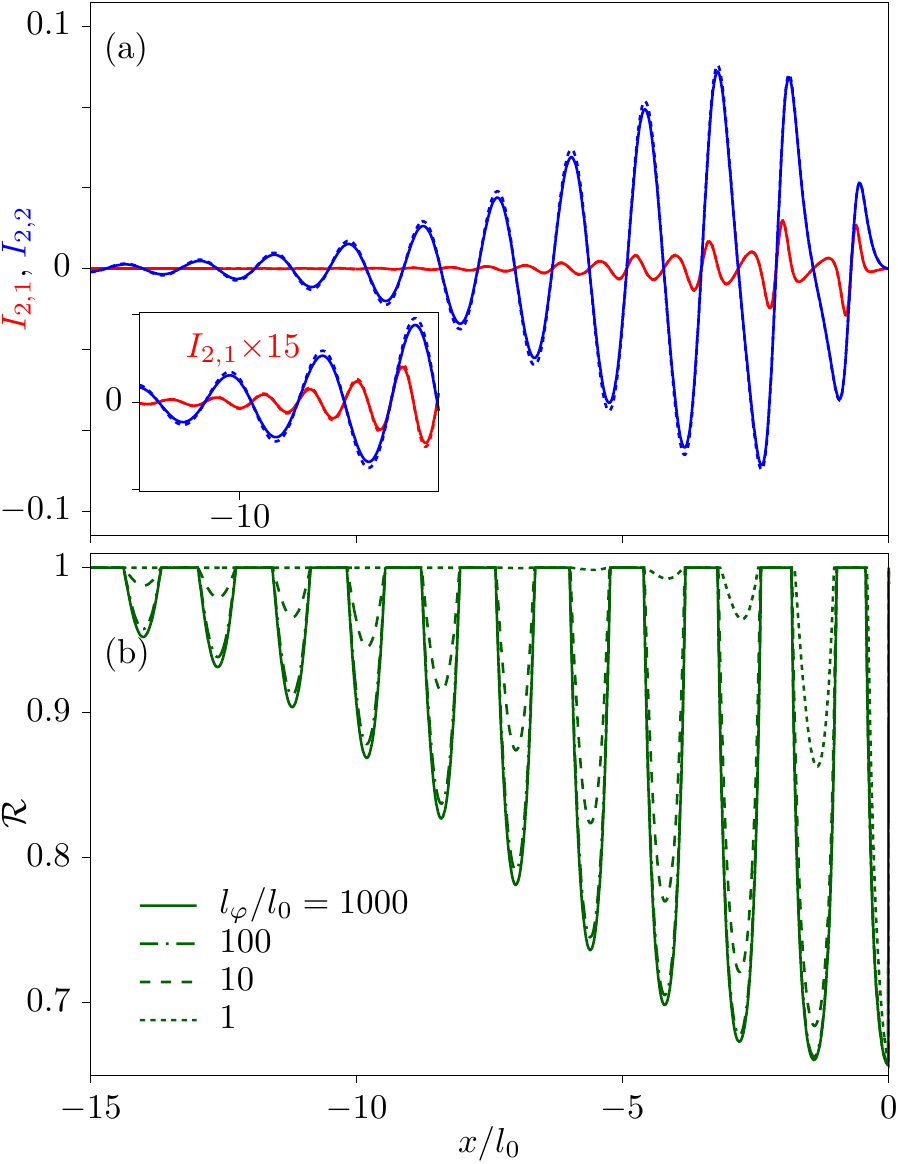}
\caption{\label{fig:longdeph}\small Effect of quasielastic scattering on the longitudinal response. (a) Longitudinal currents $I_{2,1}$ (full red line) and $I_{2,2}$ (full blue line), for $l_\varphi=10/l_0$. The case where the same $l_\varphi$ corresponds to pure dephasing is shown in dotted lines, see Fig.~\ref{fig:longpuredeph}(c) for comparison. The inset shows a zoom of some oscillations. (b) Thermoelectric rectification coefficient ${\cal R}(x)$ for different coherence lengths. All other parameters are as in Figs.~\ref{fig:muiIi} and \ref{fig:rectif}, which correspond to the case $l_\varphi\rightarrow\infty$.}
\end{figure}

Differently from the pure dephasing case of Sec.~\ref{sec:puredeph}, the probe introduces backscattering, which erases the information related to the terminal where electrons are injected. This \RS{however does not affect} the diode effect \RS{appreciably}. The asymmetry between $I_\rightarrow$ and $I_\leftarrow$ is also reduced but robust so as to still give a high rectification coefficient for strongly decoherent conductors, see Fig.~\ref{fig:longdeph}(b). However this again occurs for vanishingly small currents.

\subsection{Thermometer}
\label{sec:thermometer}

Alternatively we can use the invasive probe described by the scattering matrix $S^\Diamond$ as a thermometer~\cite{engquist_definition_1981}. For this, we assume that electrons entering the probe relax energy and thermalize due to inelastic scattering. The probe compensates this by increasing its temperature, so it does not inject any heat in the conductor. Mathematically, the electrochemical potential and temperature of the probe are adapted to the boundary conditions $I_p=J_p=0$, this time imposed to the average currents. Note the difference with the dephasing probe in Sec.~\ref{sec:dissprobe}, which imposes ${\cal I}_p(E)=0$ to the energy-resolved particle current. The distribution of the probe, rather than being out of equilibrium locally as in Eq.~\eqref{eq:f4} will be given by a Fermi distribution defined by the resulting $\mu_p$ and $T_p$. Other definitions of a local thermal probe can however be used~\cite{zhang_local_2019}.

\begin{figure}[t]
\includegraphics[width=0.95\linewidth,clip]{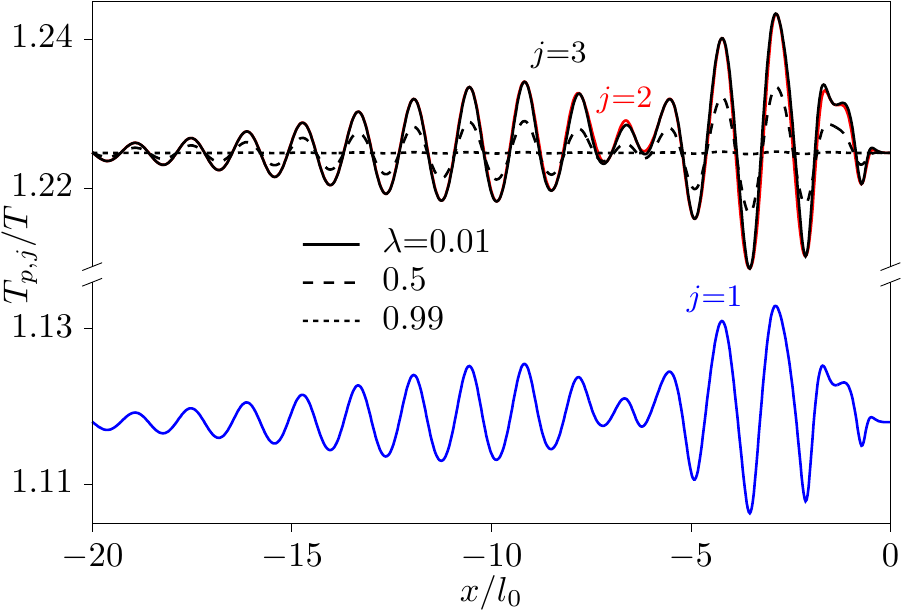}
\caption{\label{fig:tprobe}\small Temperature of the probe terminal when the temperature of terminal $j=1$ (blue), $j=2$ (red), and $j=3$ (black), is $T_j=1.5T$, with $\lambda=0.01$. Dashed and dotted lines show the case of a hot tip for increasing values of the coupling $\lambda$, as labeled. Same parameters as in Fig.~\ref{fig:muiIi}. }
\end{figure}

As a thermometer, it is important that the probe does not have its own thermoelectric response, which is achieved by considering that the coupling $\lambda$ is an energy independent constant.
Unlike the tip, this probe does not have spacial resolution, so it measures the effective temperature of the region between the tip and the scattering region~\cite{note}. 
Note that the electrochemical potential at the probe, $\mu_p$, is in general different from the one simultaneously measured by the tip, $\mu_3$.
The measured temperature is shown in Fig.~\ref{fig:tprobe}. The spacial modulation of the probe outcomes with the position of the tip shows a beating behaviour which reflects that of the heat currents flowing into the conductor terminals, see $J_1$ and $J_2$ in Fig.~\ref{fig:Ji}. The same pattern appears independently of which terminal is hot, only the amplitude of the oscillations and a global shift are affected. This makes it difficult to identify the temperature features with those of the heat current. Note that for the chosen parameters, the probe does not distinguish whether heat is injected from the tip or from terminal 2, while heat injected from terminal 1 induces a lower increase of $T_p$. This is understood in that particular case (with $\varepsilon=R=0.5$) because in the limit $\lambda\rightarrow0$, we have ${\cal T}_{p3}^{-\lambda}\approx{\cal T}_{p2}^{-\lambda}=2{\cal T}_{p1}^{-\lambda}$, see Eqs.~\eqref{eq:si4}.

Increasing the coupling to the probe, the generated currents $I_{2,j}$ are suppressed, as expected because of phase randomization in the probe (not shown). The amplitude of the temperature oscillations also decreases when $\lambda$ increases. The probe becomes less sensitive to the \RS{details of the} conductor heat currents, emphasizing the need for a weak system-probe coupling to have a meaningful measurement outcome. \RS{For large coupling, $\lambda$, the probe temperature, $T_{p,j}$ tends to the asymptotic value corresponding to $x\rightarrow0$ and $x\rightarrow\infty$.}

\section{Conclusions}
\label{sec:conclusions}

We have explored the mechanisms of thermoelectric current generation in a quantum coherent conductor locally coupled to a probe reservoir. For this, we invoked a minimal model including an ideal conductor hosting a single pointlike scattering region, in contact with a tip consisting of a single-channel splitter. The conductor lacks any intrinsic thermoelectric response, since its transmission probability is energy independent.  However, when coupling to the tip, electron-hole symmetry is broken by the quantum interference of trajectories multiply reflected between the barrier and the tip, provided the conductor is noisy (i.e., partly open). This generalizes the mechanisms involved in resonant tunneling~\cite{humphrey:02,paulsson:2003,nakpathomkun:2010,cui_peltier_2017} to multiterminal configurations. 

The generated current shows distinct properties depending on which terminal acts as the heat source. When the tip is hot, a nonlocal current is generated in the conductor. The necessary broken mirror symmetry~\cite{benenti:2017} is introduced by the position of the tip relative to the scatterer.
Spacial oscillations appear as the tip scans the conductor, and are antisymmetric when exchanging its position with respect to the scatterer. When one of the conductor terminals is heated, a longitudinal response appears. The number of oscillation nodes is doubled when the tip is between the hot terminal and the scatterer, resulting in rectifying configurations: the current flows in the same direction for opposite temperature gradients. This effect can be used to define an ideal thermoelectric diode.

The effects of phase and momentum randomization are disentangled by introducing fictitious probes. Pure dephasing is shown to suppress the amplitude of the current oscillations for tip distances longer than the dephasing length, emphasizing their interference origin. Momentum randomization by quasielastic scattering furthermore affects the mirror asymmetric properties of the current, \RS{nevertheless not affecting} the rectification coefficient \RS{in a distinct manner}. However, high rectification coefficients persist in the presence of strong dephasing. A probe inducing inelastic scattering is used to track the energy dissipation in the conductor.

We considered a simple noninteracting model which can be solved analytically. Extensions of our work to more realistic scatterers (including an intrinsic thermoelectric response, for instance) or other kinds of tips (like scanning gates that do not exchange charge with the system at all~\cite{gorini2013, steinacher_scanning_2018}), 
as well as the inclusion of electron-electron interactions~\cite{blanter_interaction_1998,imura_conductance_2002,crepieux_electron_2003,
guigou_screeningI_2009,guigou_screeningII_2009} remain as topics to be addressed in the future.

{\it Note added in proof---} We stress that we discuss effects independent of mechanisms appearing beyond linear response, such as charge accumulation~\cite{whitney_nonlinear_2013,sanchez_scattering_2013,meair_scattering_2013} and spontaneous symmetry breaking~\cite{marchegiani_nonlinear_2020}.

\RS{{\it Erratum---} There is a missing factor in the expressions for ${\cal T}_{1p}^{-\lambda}$ and ${\cal T}_{3p}^{-\lambda}$ in Eq.~\eqref{eq:si4} of the published version of this manuscript. The magenta fonts in this manuscript mark the necessary changes involving the invasive probe. The results shown in Figs.~\ref{fig:crossdeph}, \ref{fig:longdeph}, and \ref{fig:tprobe} are affected and have been corrected in this version.}

\acknowledgements
We acknowledge the QTC2017 conference in Espoo and the DPG Spring Meeting of the Condensed Matter division in Berlin 2018 where this project started taking form out of discussions that would have hardly happened via online meetings.
R.S. acknowledges funding from the Ram\'on y Cajal program RYC-2016-20778, and the Spanish Ministerio de Ciencia e Innovaci\'on via Grant No. PID2019-110125GB-I00 and through the ``Mar\'{i}a de Maeztu'' Programme for Units of Excellence in R{\&}D CEX2018-000805-M.
C.G. thanks the STherQO members for stimulating discussions.

\appendix

\section{Scattering matrices in series} \label{sec:series}

\begin{figure}[t]
\includegraphics[width=\linewidth,clip]{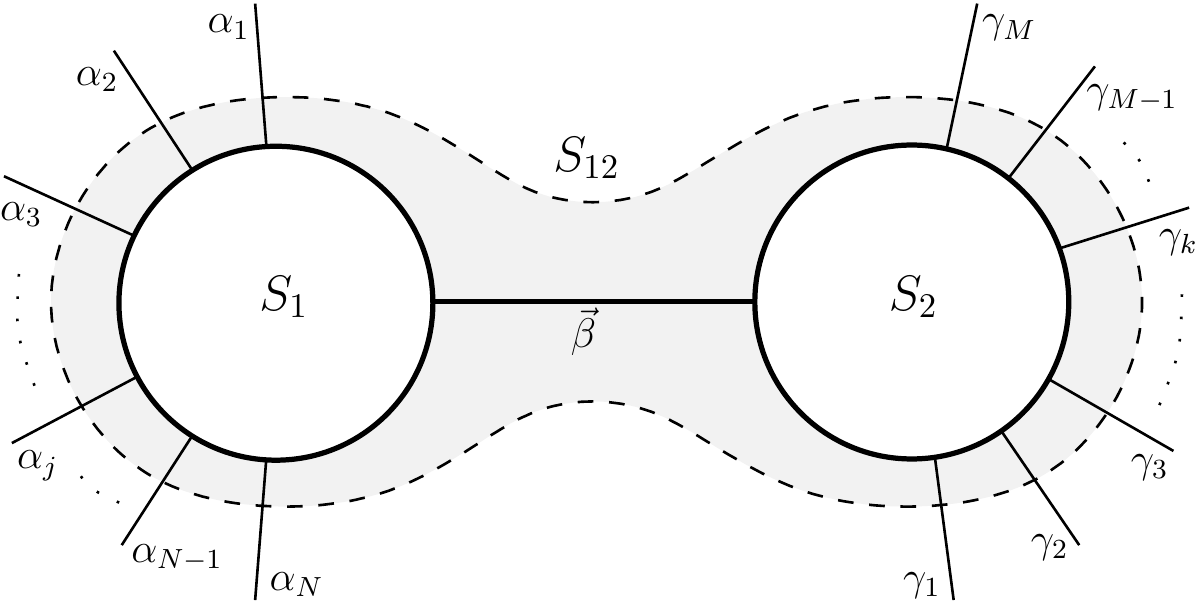}
\caption{\label{fig:comp}\small Scheme of two scattering matrices, $S_1$ and $S_2$, coupled in series, which combine to give $S_{12}$. }
\end{figure}

Consider two scattering regions $i$=1,2 which are connected in series by one or more internal channels labelled by $\vec{\beta}$, as sketched in Fig.~\ref{fig:comp}. Their respective scattering matrices can be written as
\begin{eqnarray}\label{Spart}
\displaystyle
{
\left(\begin{array}{c}
\vec{\alpha}' \\ 
\vec{\beta}' \\
\end{array}  \right)
=S_1
\left(\begin{array}{c}
\vec{\alpha} \\ 
\vec{\beta} \\
\end{array}  \right)
\quad \text{and}\quad 
\left(\begin{array}{c}
\vec{\beta} \\
\vec{\gamma}' \\ 
\end{array}  \right)
=S_2
\left(\begin{array}{c}
\vec{\beta}' \\
\vec{\gamma} \\ 
\end{array}  \right)\nonumber
}
\end{eqnarray}
where the outgoing waves in channels $\vec{\beta}'$ of region 1 are ingoing waves of region 2 (and viceversa with $\vec{\beta}$). The scattering matrices are:
\begin{eqnarray}\label{Si}
\displaystyle
{
S_i=
\left(\begin{array}{cc}
r_i & t'_i \\ 
t_i & r'_i\\
\end{array}  \right).
}
\end{eqnarray}
The scattering matrix of the whole system 
\begin{eqnarray}\label{Stot}
\displaystyle
{
\left(\begin{array}{c}
\vec{\alpha}' \\ 
\vec{\gamma}' \\
\end{array}  \right)
=S_{12}
\left(\begin{array}{c}
\vec{\alpha} \\ 
\vec{\gamma} \\ 
\end{array}  \right),
\quad\text{with}\quad
S_{12}=
\left(\begin{array}{cc}
r & t' \\ 
t & r'\\
\end{array}  \right),
\nonumber
}
\end{eqnarray}
is obtained by simple linear algebra, leading to 
\begin{align}
r&=r_1+t'_1r_2(1-r_1'r_2)^{-1}t_1\\
t&=t_2(1-r_1'r_2)^{-1}t_1\\
r'&=r_2'+t_2(1-r_1'r_2)^{-1}r'_1t'_2\\
t'&=t'_1\left[1+r_2(1-r_1'r_2)^{-1}r'_1\right]t'_2.
\end{align}

In the previous, we have ignored the phases accumulated along the connecting channels. Usually these are the only phases with a physical meaning.  A convenient way to take them into account is by taking all arbitrary phases in $S_1$ and $S_2$ to zero and treating the propagation along the connecting channels as an additional scattering region for which waves are perfectly transmitted along every channel $i$ gaining a phase $\phi_i$. For a single channel connection, and in the absence of a magnetic field, this can be written as 
\begin{eqnarray}\label{Sph}
\displaystyle
{
S_{\rm ph}(\varphi)=
\left(\begin{array}{cc}
0 & e^{i\varphi} \\ 
e^{i\varphi} & 0\\
\end{array}  \right).
}
\end{eqnarray}
In the case of having $N$ channels, we have
\begin{equation}
S_{\rm ph}(\vec{\varphi})=\sum_{i=1}^NS_{\rm ph}(\varphi_i)\otimes{\rm diag}_N(\delta_{ii}),
\end{equation}
where ${\rm diag}_N(\delta_{ii})$ is a $N\times N$ matrix whose only nonvanishing element is the $i$th element of the diagonal, which is 1.
The total scattering matrix is hence the result of combining the three matrices $S_1$, $S_{\rm ph}(\vec{\varphi})$, and $S_2$.

\section{Linear coefficients for heat currents}
\label{app:linheat}
In the linear regime, assuming the probe condition for the tip, $I_3=0$, we get for the heat currents
\beq
J_i={\cal M}_i(\mu_2-\mu_1)+\sum_j{\cal K}_{ij}\delta T_j,
\eeq
with ${\cal M}_1=-{\cal M}_2-{\cal M}_3$,
\begin{align}
{\cal M}_2&=M_{21}+\frac{G_{13}M_{23}}{G_{13}+G_{23}}\\
{\cal M}_3&=\frac{G_{32}M_{31}-G_{13}M_{23}}{G_{13}+G_{23}},
\end{align}
and the thermal conductances:
\beq
{\cal K}_{ij}=q_{\rm H}\delta_{ij}-K_{ij}-\frac{M_{i3}L_{3j}}{G_{31}+G_{32}}.
\eeq

\section{Limitations of the Sommerfeld expansion} \label{sec:sommlim}

\begin{figure}[t]
\includegraphics[width=\linewidth,clip]{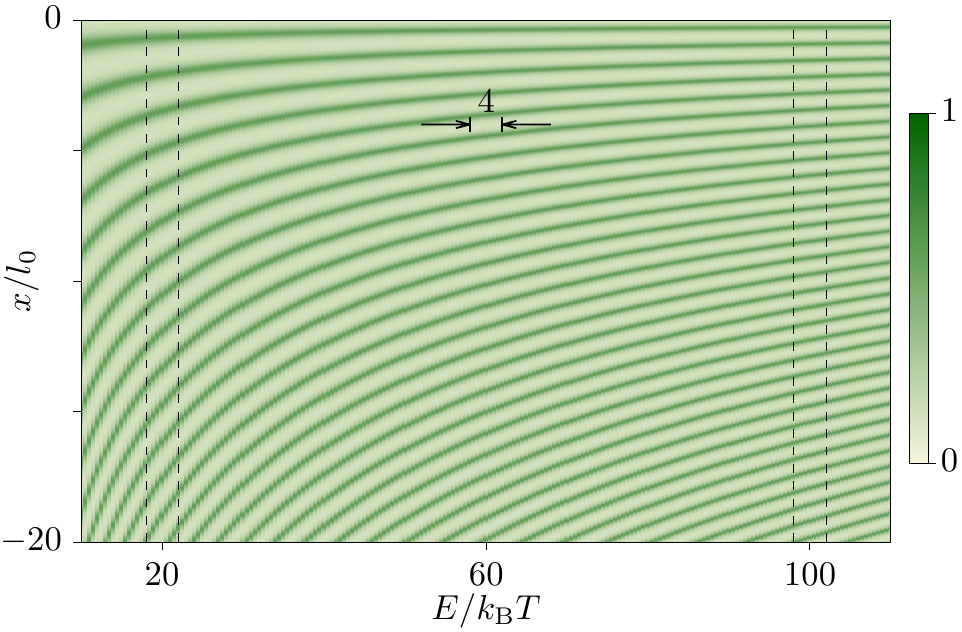}
\caption{\label{fig:tEx}\small Energy dependence of the transmission probability ${\cal T}_{23}$ as a function of the tip position. The relevant temperature scale $4\kBT$ around the chemical potentials $\mu=20\kBT$ (considered in most figures) and $\mu=100\kBT$ (considered in one of the cases in Fig.~\ref{fig:thermcurrU}) is delimited by dashed lines as a guide to the eye.}
\end{figure}

The Sommerfeld expansion is known to behave well for low temperatures $\kBT\ll\mu$, provided the transmission probabilities are smooth. In our system, the oscillatory energy dependence of the transmission probabilities sets an additional limitation depending on the position of the tip at finite temperatures. The Sommerfeld approximation breaks down when the energy associated to the distance of two peaks, 
\beq
\Delta E=\frac{2\pi^2\hbar^2}{mx^2}\left(1-\frac{1}{\pi}k|x|\right),
\eeq 
becomes comparable to the thermal energy. This depends on the chemical potential around which the current integrals are performed. At short distances, $\Delta E$ diverges, as shown in Fig.~\ref{fig:tEx} for the case of ${\cal T}_{23}$, which stays constant over all energies as $x\rightarrow0$ (similar behaviour is obtained for ${\cal T}_{12}$ and ${\cal T}_{13}$). $\Delta E$ decreases when the tip separates from the scatterer, most dramatically for low energies. 

\section{Coupling to a dephasing probe}
\label{sec:probabpd}

We obtain the scattering matrix for pure dephasing processes by assuming that the coupling to each of the fictitious probes, $a$ and $b$, is defined by a tunnel barrier with transmission probability $\tau$, cf. Fig.~\ref{fig:Sdeph}(a). The coupling is the same for both probes, so no propagation direction is privileged. We can simply combine the elements of two scattering matrices of the form of Eq.~\eqref{Ssq}, choosing $\phi=0$ in order to avoid the dephasing probe to introduce unnecessary phases. 
With this, we arrive at
\begin{eqnarray}\label{Sdeph}
\displaystyle
{
S^\ocircle=
\left(\begin{array}{cccc}
0 & \sqrt{1{-}\tau} & 0 & i\sqrt{\tau}\\ 
\sqrt{1{-}\tau} & 0 & i\sqrt{\tau} & 0\\
i\sqrt{\tau} & 0 & \sqrt{1{-}\tau} & 0\\
0 & i\sqrt{\tau} & 0 & \sqrt{1{-}\tau}\\
\end{array}  \right).
}
\end{eqnarray}
Note that different conventions are also possible~\cite{Datta1995}.

We now calculate the transmission probabilities, modified by the presence of the probes, ${\cal T}_{\alpha\beta}^{\pm\tau}$. The indices $\alpha,\beta$ accounting for all conductor terminals (indices $i$ and $j$ will still be used when the probes are explicitly excluded).
The transmission probabilities, irrespective of the relative position of the tip, ${\cal T}_{\alpha\beta}^\tau$ are obtained from ${\cal T}_{\alpha\beta}^{\pm\tau}$ as it was done in Eq.~\eqref{eq:Tij}.
Note that they do not depend on the position of the probes, but only on how strongly they are coupled to the conductor. 
In the strongly coupled limit with $\tau=1$, we get ${\cal A}_\tau=1$ and $\zeta_\tau=0$, i.e. every phase dependence is lost.

Considering that the tip position is $x<0$, the transmission probabilities between the conductor terminals and the tip are modified by the coupling to the pure dephasing probes ($S^\ocircle$) with respect to what we got in Eq.~\eqref{eq:transm}:
\begin{align}
\label{eq:sijpd}
{\cal T}_{12}^{-\tau}&=\frac{(1-R)(1-\tau)(\eta_+-\varepsilon)}{2{\cal A}_{\tau}}\nonumber\\
{\cal T}_{13}^{-\tau}&=\frac{\varepsilon({\cal A}_\tau+\zeta_\tau)}{{\cal A}_\tau}\\
{\cal T}_{23}^{-\tau}&=\frac{(1-R)(1-\tau)\varepsilon}{{\cal A}_\tau},\nonumber
\end{align}
where ${\cal A}_\tau$ and $\zeta_\tau$ are respectively obtained from ${\cal A}$ and $\zeta$ [Eqs.~\eqref{eq:cA} and \eqref{eq:z}] by replacing $R\rightarrow R_\tau=R(1-\tau)^2$. They are now conditioned on the electrons being reflected at the coupling to the probes.

The probabilities for an electron injected from a conductor terminal to be absorbed by one of the probes are:
\begin{align}
\label{eq:Tiab}
{\cal T}_{a1}^{-\tau}&=\frac{\tau\eta_+^2}{4{\cal A}_\tau}&
{\cal T}_{b1}^{-\tau}&=(1{-}\tau)R{\cal T}_{a1}^{-\tau}\nonumber\\
{\cal T}_{b2}^{-\tau}&=\frac{\tau(1-R)}{{\cal A}_\tau}&
{\cal T}_{a2}^{-\tau}&=(1{-}\tau)\eta_-^2{\cal T}_{b2}^{-\tau}/4\\
{\cal T}_{a3}^{-\tau}&=\frac{\varepsilon\tau}{{\cal A}_\tau}&
{\cal T}_{b3}^{-\tau}&=(1{-}\tau)R{\cal T}_{a3}^{-\tau}.\nonumber
\end{align}
For electrons to be transmitted from one probe to the other we find
\begin{align}
\label{eq:Tiab}
{\cal T}_{ab}^{-\tau}=\frac{\tau^2\eta_-^2}{4{\cal A}_\tau}\quad {\rm and}\quad
{\cal T}_{ba}^{-\tau}=\frac{\tau^2R}{{\cal A}_\tau}.
\end{align}
By symmetry we obtain the remaining ones: ${\cal T}_{ai}^{-\tau}={\cal T}_{ib}^{-\tau}$ and ${\cal T}_{bi}^{-\tau}={\cal T}_{ia}^{-\tau}$, and, for the internal reflection at the probes: ${\cal T}_{aa}^{-\tau}=1-\sum_{\alpha\neq a}{\cal T}_{a\alpha}^{-\tau}$ and ${\cal T}_{bb}^{-\tau}=1-\sum_{\alpha\neq b}{\cal T}_{b\alpha}^{-\tau}$.

The probabilities ${\cal T}_{\alpha\beta}^{+\tau}$ are obtained from the previous expressions for ${\cal T}_{\alpha\beta}^{-\tau}$, by replacing $1\leftrightarrow2$ and $a\leftrightarrow b$.

The probe conditions ${\cal I}_a(E)={\cal I}_b(E)=0$ are satisfied only if the probes acquire a nonequilibrium distribution, in particular
\beq
\label{eq:purdephdistr}
f_\alpha=\sum_iY_{\alpha i}f_i,
\eeq
for terminals $\alpha=a,b$, with $Y_{\alpha i}=y_{\alpha i}/\sum_iy_{\alpha i}$ and 
\beq
y_{a i}=(1-{\cal T}_{aa}^{\tau}){\cal T}_{a i}^{\tau}+{\cal T}_{ab}^{\tau}{\cal T}_{b i}^{\tau},
\eeq
and likely for $y_{bi}$ by replacing $a\leftrightarrow b$. Note that equilibrium properties are recovered as $\sum_iY_{ai}=\sum_iY_{bi}=1$.

The currents in the conductor terminals can then be calculated from
\beq
{\cal I}_i=\frac{2e}{h}\sum_{i=1}^3\left(\delta_{ij}-\tilde{\cal T}_{ij}\right)f_i(E),
\eeq
with the generalized transmissions:
\beq
\tilde{\cal T}_{ij}={\cal T}_{ij}^\tau+{\cal T}_{ia}^\tau Y_{aj}+{\cal T}_{ib}^\tau Y_{bj}.
\eeq

\section{Coupling to an invasive probe}
\label{app:sdeph}

Consider that the conductor is coupled to a fictitious probe $p$ via two additional channels, as shown in Fig.~\ref{fig:Sdeph}(b). Imposing that the different channels are not reflected at the probe (i.e., the diagonal elements of the scattering matrix are zero) and that the probe couples symmetrically to the conductor, one finds that the scattering matrix is given by $S_{ij}^\Diamond=\sigma_{ij}^\Diamond e^{i(\theta_i+\theta_j)}$, with~\cite{buttiker:1986}
\begin{eqnarray}\label{Sdep}
\displaystyle
{
\sigma^\Diamond=
\left(\begin{array}{cccc}
0 & \sqrt{1{-}\lambda} & \sqrt{\lambda} & 0\\ 
\sqrt{1{-}\lambda} & 0 & 0 & \sqrt{\lambda}\\
\sqrt{\lambda} & 0 & 0 & -\sqrt{1{-}\lambda}\\
0 & \sqrt{\lambda} & -\sqrt{1{-}\lambda} & 0\\
\end{array}  \right),
}
\end{eqnarray}
where $\lambda$ parametrizes the conductor-probe coupling. Note however that each probe channel is coupled to only one conductor channel, so nothing avoids an electron absorbed by the probe to be reemitted into the same channel, resulting in the reversal of its momentum. For $\lambda=0$, the probe is not coupled. For $\lambda=1$, every electron incoming from the conductor channels are absorbed by the probe.
Note that, also in this case, the result does not depend on the position of the probe. 

As for the pure dephasing case in Appendix~\ref{sec:probabpd}, we need a four channel scattering matrix (two channels in the conductor and two connected to the probe): a single channel connection (as for the model of the tip) does not include the desired properties that all electrons are absorbed by a strongly coupled probe~\cite{buttiker:1986} and that it does not add  additional phases to the dynamics.

In this case, the transmission probabilities between conductor terminals ${\cal T}_{ij}^{-\lambda}$ can be obtained from Eqs.~\eqref{eq:sijpd} by simply replacing $\tau\rightarrow\lambda$. For the new coefficient ${\cal A}_\lambda$ one additionally needs to replace $\chi\rightarrow\chi_\lambda=\chi+2(\theta_1+\theta_2)$.
The transmission probabilities from the probe (considering the contribution of the two channels) are
\begin{align}
\label{eq:si4}
{\cal T}_{1p}^{-\lambda}&=\frac{\lambda(\eta_+-\varepsilon)[1+R\RS{(1-\lambda)}]}{2{\cal A}_{\lambda}}\nonumber\\
{\cal T}_{2p}^{-\lambda}&=\frac{\lambda(1-R)[1+(1-\lambda)(\eta_--\varepsilon)]}{{\cal A}_\lambda}\\
{\cal T}_{3p}^{-\lambda}&=\frac{\lambda\varepsilon[1+R\RS{(1-\lambda)}]}{{\cal A}_\lambda}.\nonumber
\end{align}
For the probabilities ${\cal T}_{\alpha\beta}^{+\lambda}$, one needs to exchange 1$\leftrightarrow$2 in the corresponding expressions for ${\cal T}_{\alpha\beta}^{-\lambda}$.
Again, all phases are lost for $\lambda=1$.

In this case, the probe maintains the symmetry ${\cal T}_{\alpha\beta}^\lambda={\cal T}_{\beta\alpha}^\lambda$.

In the case of imposing a quasi-elastic boundary condition ${\cal I}_p(E)=0$, as is done in Sec.~\ref{sec:dissprobe}, we solve for the distribution of the probe, which becomes 
\beq
\label{eq:f4}
f_p(E)=\frac{{\cal T}_{p1}^{\lambda}f_1(E)+{\cal T}_{p2}^{\lambda}f_2(E)+{\cal T}_{p3}^{\lambda}f_3(E)}{2-{\cal T}_{pp}^{\lambda}},
\eeq
where the factor 2 in the denominator accounts for the two channels of terminal $p$.
The currents at the other conductor terminals $i=1,2,3$ can then be written from
\beq
{\cal I}_i=\frac{2}{h}\sum_{j=1}^3\left({\cal T}_{ij}^{\lambda}+\frac{{\cal T}_{ip}^{\lambda}{\cal T}_{pj}^{\lambda}}{2-{\cal T}_{pp}^{\lambda}}\right)(f_i-f_j),
\eeq
where the second term in the first brackets (proportional to $\lambda$) introduces the transition from the coherent to the incoherent transport regimes~\cite{buttiker:1988}.
\bibliography{biblio.bib}

\begin{thebibliography}{136}
\bibitem [{not()}]{note}%
  \BibitemOpen
  \href@noop {} {}\bibinfo {note} {We have already discussed how a spatially
  resolved measurement (the one made by the tip) introduces privileged phases
  resulting in thermoelectric contributions. The thermometer does not yield any
  such contributions, precisely since it lacks resolution.}\BibitemShut {Stop}%
\end{thebibliography}

\end{document}